\documentstyle[aps,epsfig,floats]{revtex}

\tighten

\def\ii{{\rm i}}
\def\de{{\rm\,d}}
\def\e{{\rm e}}

\def\<{\langle} 
\def\>{\rangle} 
\def\tr{\mbox{Tr}\,}

\def\eq{\begin{equation}}
\def\feq{\end{equation}}
\def\be{\begin{equation}}
\def\ee{\end{equation}}
\def\ar{\begin{eqnarray}}
\def\far{\end{eqnarray}}
\def\bea{\begin{eqnarray}}
\def\eea{\end{eqnarray}}
\def\^{\hat}

\newcommand{\bm}[1]{\mbox{\boldmath $#1$}}
\newcommand{\g}{\gamma}
\newcommand{\sig}{\sigma}
\newcommand{\eps}{\epsilon}
\newcommand{\nn}{\nonumber}

\newcommand{\Pslash}{\kern 0.2 em P\kern -0.56em \raisebox{0.3ex}{/}}
\newcommand{\Sslash}{\kern 0.2 em S\kern -0.56em \raisebox{0.3ex}{/}}
\newcommand{\pslash}{\kern 0.2 em p\kern -0.4em /}
\newcommand{\kslash}{\kern 0.2 em k\kern -0.45em /}
\newcommand{\qslash}{\kern 0.2 em q\kern -0.4em /}
\newcommand{\nslash}{\kern 0.2 em n\kern -0.5em /}
\newcommand{\one}{1\hspace{-1.9pt}\rule[0.08ex]{0.45pt}{1.5ex}\hspace{2pt}}
\newcommand{\xbj}{x_{\scriptscriptstyle B}}                 

\def\bold#1{\setbox0=\hbox{$#1$}%
     \kern-.02em\copy0\kern-\wd0
     \kern.04em\copy0\kern-\wd0
     \kern-.02em\raise.0433em\box0 }

\begin{document}
 
\title{
\begin{flushright}
\begin{minipage}{4 cm}
\small
VUTH 00-20\\
\end{minipage}
\end{flushright}
Deep inelastic leptoproduction of spin-one hadrons}

\author{A.~Bacchetta$^1$, P.J.~Mulders$^1$}

\address{
$^1$Division of Physics and Astronomy, Faculty of Science, Free University \\
De Boelelaan 1081, NL-1081 HV Amsterdam, the Netherlands\\[2mm]}

\date{July 12, 2000}
\maketitle
\begin{abstract}
In this paper we analyze deep inelastic one-particle 
inclusive processes for the case of spin-one
targets or for the case of spin-one produced hadrons, 
such as $\rho$ mesons.  This allows the measurement of
new distribution and fragmentation functions not available in 
the spin-half case, and provides new ways to probe functions otherwise
difficult to measure. We will analyze only contributions leading order 
in $1/Q$, 
but we will include effects of the transverse momentum of partons.
We also include time-reversal odd functions.
\end{abstract}

\section{Introduction}

Cross-sections in deep inelastic scattering (DIS) can be expressed
in terms of distribution and fragmentation functions, which provide 
information on the quark and gluon structure of hadrons.
The energy scale of the process is given by $Q^2= -q^2$, 
$q$ being the four-momentum transfer of the lepton. Depending on
the number of observables one is able to measure, one can extract 
a variety of functions. 
The functions appearing in leading order in $1/Q$ can be interpreted as
partonic probability densities.

We will study the case of one-particle inclusive experiments, which require 
measurement of one hadron among the ones produced 
in the scattering event. 
We will emphasize the importance of including transverse momenta
of partons. We will also include T-odd functions.
We will give a systematic list of the various functions
that come into play up to leading order in $1/Q$  
when we deal with either spin-1 targets or spin-1 outgoing hadrons. 
The second case is of interest to analyze vector meson production. 

To properly study the distribution and fragmentation functions including
transverse momentum dependence, we
will  start from a field-theoretical formalism, as outlined in \cite{multa}. 
This approach has been fully exploited only to study spin-$\frac{1}{2}$
targets and spin-$\frac{1}{2}$ outgoing hadrons.
After an overview of the general properties of spin-1 particles and of the 
general formalism needed to deal with them (Sec.~\ref{s:spinone}), 
we turn to the most general parameterization of the correlation functions 
when spin-1 is 
included and we define the distribution and fragmentation functions
(Sec.~\ref{s:main}). Distribution and fragmentation functions 
integrated over transverse momenta have been partially studied already
in a number of papers~\cite{hjm,ji,sst}. An incomplete treatment of transverse
momentum dependent functions has been performed in~\cite{hk}.

The distribution functions for a spin-1 target could be used for the
deuteron, but is not the main goal of our study as the deuteron is in
essence a weakly bound system of two spin-$\frac{1}{2}$ nucleons. 
The spin-1 distribution functions are useful as a passage 
to the fragmentation functions for spin-1 hadrons. 
The latter, however, require final state polarimetry of the produced hadron, 
i.e. the study of the angular distribution of its decay products. 
The most common of such hadrons is the $\rho$ meson. It is abundantly 
produced in leptoproduction experiments, e.g. at HERA, and it should be 
possible
to measure its polarization in a detailed, as it has already been done
in diffractive production \cite{zeu,h1,her} and in hadronic Z$^0$ decay
\cite{delphi}.  
Another possibility is observation of polarization in
inclusive leptoproduction of $\phi$ mesons, for which there should be less
 hadronic background.
   
In the last section we focus more specifically on deep-inelastic
leptoproduction of spin-1 hadrons and we list all the possible cross-sections 
for different polarization conditions in terms of the usual spin-$\frac{1}{2}$
distribution functions and the newly defined spin-1 fragmentation functions. 

\section{The description of spin-one particles}
\label{s:spinone}

The description of particles with spin can be attained by using a spin density
matrix ${\bm \rho}$ in the rest-frame of the particle.
The parameterization of the density matrix for a spin-J particle is
conveniently  performed with the introduction of irreducible spin tensors up
to rank 2J.
For example, the density matrix of a spin-$\frac{1}{2}$ particle can be 
decomposed on a Cartesian basis of $2 \times 2$ matrices, 
formed by the identity matrix and the three Pauli matrices,
\eq
{\bm \rho}=\frac{1}{2}\left({\bm 1} + S^i {\bm\sigma}^i \right),
\feq
where we introduced the (rank-one) spin vector $S^i$.

To parameterize the density matrix of a spin-1 particle we can choose a 
Cartesian basis of
$3\times 3$ matrices, formed by the identity matrix,
three spin matrices ${\bm\Sigma}^i$ (generalization of the Pauli matrices 
to the three-dimensional case) and five extra matrices ${\bm\Sigma}^{ij}$. 
These last ones
can be built using bilinear combinations of the spin matrices. In three 
dimensions these combinations are no more dependent on the spin matrices 
themselves, as it would be for the Pauli matrices.
We choose them to be (see \cite{bls} and \cite{mad} for a comparison)
\eq
{\bm\Sigma}^{ij}=\frac{1}{2}\left({\bm\Sigma}^i {\bm\Sigma}^j 
	+{\bm\Sigma}^j {\bm\Sigma}^i\right)
		-\frac{2}{3}{\bf 1} \; \delta^{ij}.
\feq

With these preliminaries, we can write the spin density matrix as
\eq
{\bm \rho}=\frac{1}{3}\left({\bf 1} + \frac{3}{2} S^i {\bm\Sigma}^i 
		+ 3\, T^{ij} {\bm\Sigma}^{ij}\right),
\label{e:density}
\feq
where we introduced the symmetric traceless rank-two spin tensor $T^{ij}$.

We choose the following  way of parameterizing the spin vector and tensor
in the rest-frame of the hadron,
\begin{eqnarray}
{\bm S}&=&\left(S_{T}^x, S_{T}^y, S_{L}\right),    \label{e:vector}\\
&&\nonumber \\
{\bm T}&=&\frac{1}{2}\left(\begin{array}{ccc}
	-\frac{2}{3}{S_{LL}}+{S_{TT}^{xx}} 
		& {S_{TT}^{xy}}	& {S_{LT}^{x}} \\
	{S_{TT}^{xy}}& 	-\frac{2}{3}{S_{LL}}-{S_{TT}^{xx}}
					& {S_{LT}^{y}}  \\
	 {S_{LT}^{x}}	& {S_{LT}^{y}}	& \frac{4}{3}{S_{LL}} \\
	\end{array}\right).
\label{e:tensor}
\end{eqnarray} 
In App.~\ref{a:tensor} we give some explicit forms and other
details of the density matrices and parameters involved 
in Eq.~(\ref{e:tensor}).
In an arbitrary frame, different from the rest-frame, the spin 
vector and tensor satisfy
the conditions $P_{\mu}S^{\mu}=0$ and $P_{\mu}T^{\mu \nu}=0$,
where $P_{\mu}$ is the momentum of the hadron. 
In App.~\ref{a:decay} we also discuss how the tensor polarization
of a produced $\rho$-meson  can be extracted from the angular distribution
of the decay products $\pi^+ \pi^-$.

\section{Correlation functions}
\label{s:main}

Cross sections of DIS events are proportional to the contraction between
a purely leptonic tensor and a purely hadronic tensor. While the leptonic 
tensor can be calculated theoretically, we are not able to do the same for 
the hadronic tensor, because
we lack knowledge of the inner, non-perturbative structure of hadrons.

In the Bjorken limit, it is possible to separate the hadronic tensor
into a hard part (virtual photon-quark scattering) and a 
soft part, containing the information on the parton distribution inside 
the hadron. This soft part is a correlation function, defined as the
matrix element of quark fields between hadronic states. 
In one-particle inclusive processes we need two correlation functions, one
describing the quark distributions in the target hadron and one describing the
hadronization of a quark into the detected final state hadron.  

\begin{figure}
        \centering
        \epsfig{figure=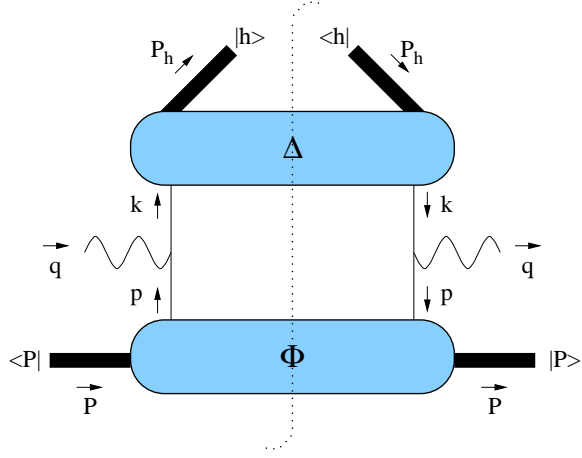,height=6cm}
        \caption{Diagrammatic representation of semi-inclusive DIS}
        \label{f:bull}
        \end{figure}

In leading order in $1/Q$ (also referred to as ``leading twist'' or
``twist-2'') we are  concerned only with 
quark-quark correlation functions entering the handbag diagram in
Fig.~\ref{f:bull}. They are
defined as follows (using Dirac indices $\alpha$ and $\beta$):
\ar
\Phi_{\alpha \beta}(p,P,S,T)
&=& \int \frac{\de^{4}\xi}{(2\pi)^{4}}
                \e^{- \ii p \cdot \xi}
        \<P,S,T|\overline{\psi}_{\beta}(\xi)\psi_{\alpha}(0)|P,S,T\>,
\label{e:corrphi}\\
\Delta_{\alpha\beta}(k,P_h,S_h,T_h)&=&\int
        \frac{\de^{4}\xi}{(2\pi)^{4}}\e^{+\ii k \cdot \xi}
        \<0|\psi_{\alpha}(\xi)|P_h,S_h,T_h\>\<P_h,S_h,T_h|
             \overline{\psi}_{\beta}(0)|0\>,   
\label{e:corrdelta}
\far
and describe the quark distribution and fragmentation, respectively.
Here, $p$ is the momentum of the quark emerging from the target, while $k$ is 
the momentum of the quark decaying into an outgoing hadron after being struck
by a virtual photon (see Fig.~\ref{f:bull}). The vector $P$ ($P_h$) is the
momentum of the hadronic target (outgoing hadron), the quantities $S$ ($S_h$)
and  $T$ ($T_h$) are the spin vector
and tensor. 

The correlation functions can be expressed in several terms, each one being a
combination of the Lorentz vectors $p$ ($k$) and $P$ ($P_h$), the Lorentz
pseudo-vector $S$ ($S_h$), the Lorentz tensor $T$ ($T_h$) and the Dirac 
structures
\[	
\one,\; \g_5,\; \g^{\mu},\; \g^{\mu}\g_5,
	\;\ii \sig^{\mu \nu}\g_5.
\]
The spin vector and tensor can only appear linearly in the 
decomposition. 
Moreover, each term of the full expression has to fulfill the conditions of 
hermiticity and parity invariance
\ar
\Phi(p,P,S,T)& =&\makebox[5cm][l]{$\g^0 \Phi^{\dag}(p,P,S,T) \g^0$}
	\mbox{hermiticity,}
\label{e:herm}\\             
\Phi(p,P,S,T)& =&\makebox[5cm][l]
	{$ \g^0 \Phi(\tilde{p},\tilde{P},-\tilde{S},\tilde{T}) \g^0$}
	\mbox{parity,}
\label{e:parity}
\far
where $\tilde{p}$, $\tilde{P}$ and $\tilde{S}$ represent respectively the 
vectors $p$, $P$ and $S$ having space components with inverted sign and
$\tilde{T}$ represents the tensor $T$ having mixed space-time components with
inverted sign.
For the distribution part $\Phi$ one also obtains a constraint from
time-reversal inversion (leaving out effective T-odd parts coming from
for instance gluonic poles~\cite{hts,bmt})

\ar
\Phi(p,P,S,T) & =&\makebox[5cm][l]
	{$\g^1 \g^3 \, \Phi^* (\tilde{p},\tilde{P},\tilde{S},\tilde{T})  
			\,\g^3 \g^1  $}
	\mbox{time-reversal}.
\far 

For the fragmentation part $\Delta$, containing out-states in the definition,
time-reversal invariance cannot be used as a constraint~\cite{rkr,hhk,jj} and
one is left with the so-called time-reversal odd (T-odd) contributions,
leading in particular to interesting single spin asymmetries~\cite{col,boe}.  
We will include the T-odd contributions in our discussion for $\Phi$,
because it will be used as the general case of correlation functions.
Throughout the rest of the article we will put
time-reversal odd terms between brackets to make them easily identifiable.

The most general decompositions of the correlation function $\Phi$ imposing
hermiticity and parity is
\ar 
\Phi(p,P,S,T)  &=& 
                M\,A_1\,\one + A_2\,\Pslash + A_3\pslash
                +\left(\frac{A_{4}}{M}\,\sigma_{\mu \nu} P^{\mu} p^{\nu}\right)
                +\left( \ii A_5\;p\cdot S\, \gamma_5\right)
                                \nonumber \\
  & &		\mbox{}	+ M\,A_6 \,\Sslash\, \gamma_5
                + A_7\,\frac{p\cdot S}{M}\,\Pslash\, \gamma_5     
                + A_8\,\frac{p\cdot S}{M} \pslash\, \gamma_5
                + \ii A_9\,\sig_{\mu\nu}\gamma_5 S^{\mu}P^{\nu}
                              \nn \\
  & &           \mbox{}+ \ii A_{10}\,\sig_{\mu\nu}\gamma_5 S^{\mu}p^{\nu}
    +\ii A_{11}\,\frac{p\cdot S}{M^2}\,\sig_{\mu\nu}\gamma_5 P^{\mu}p^{\nu}
	    +\left( A_{12}\, \frac{\eps_{\mu \nu \rho \sigma}\gamma^\mu P^\nu
              p^\rho S^\sigma}{M}\right)                     \nn \\
  & &		\mbox{}+\frac{A_{13}}{M}\,p_{\mu}p_{\nu}T^{\mu \nu}\,\one
		+\frac{A_{14}}{M^2}\;p_{\mu}p_{\nu}T^{\mu \nu}\Pslash
		+\frac{A_{15}}{M^2}\;p_{\mu}p_{\nu}T^{\mu \nu}\pslash \nn \\
  & &		\mbox{}+\left(\frac{A_{16}}{M^3}\;p_{\mu}p_{\nu}T^{\mu \nu}
			\sig_{\rho \sigma} P^{\rho}p^{\sig}\right) 
    + A_{17}\;p_{\mu}T^{\mu \nu}\g_{\nu}
    +\left(\frac{A_{18}}{M}\,\sig_{\nu \rho}P^{\rho}\,p_{\mu}T^{\mu\nu}\right)
		\nn \\
  & &		\mbox{}
    +\left(\frac{A_{19}}{M}\,\sig_{\nu \rho}p^{\rho}\,p_{\mu}T^{\mu\nu}\right)
    +\left(\frac{A_{20}}{M^2}\,\eps_{\mu \nu \rho \sigma} 
		\g^{\mu} \g_5 P^{\nu} p^{\rho}\, p_{\tau}T^{\tau \sig} \right).
					\label{e:decomphi}  
\far

The amplitudes $A_i$ are real functions $A_i=A_i(p\cdot P,p^2)$. The 
decomposition of the correlation function $\Delta$ is analogous. The 
amplitudes $A_4$, $A_5$, $A_{12}$, $A_{16}$, $A_{18}$, $A_{19}$ and
$A_{20}$ are T-odd.

In order to select leading twist contributions we
perform a Sudakov decomposition of the Lorentz structures we have. 
We  choose two light-like vectors $n_+$ and $n_-$ satisfying $n_+\cdot
n_-=1$. We will call the plane perpendicular to these vectors ``transverse
plane''. We define the two projectors
\begin{eqnarray} 
g_T^{\mu \nu}&=& g^{\mu\nu}
	-n_+^{\{ \mu}n_-^{\nu\}}, \label{e:gtrans}\\
\eps_T^{\mu \nu}&=& \eps^{\mu \nu \rho \sig}n_{+\, \rho}n_{-\,\sig},
\end{eqnarray}
where the curly braces around the indices denote symmetrization of these
indices. 
Given a vector $a^{\mu}$ we will sometimes make use of the notation 
$a_T^{\mu}=g_T^{\mu \nu}a_{\nu}$ and we will denote its
two-dimensional component lying in the transverse plane as 
${\bm a}_T$.

We assume the following decompositions of the Lorentz structures we are 
interested in:
\ar
P^{\mu} &=& 
P^+ n_+^{\mu}
+ \frac{M^2}{2P^+}n_-^{\mu} , 
\label{e:decomp}\\
p^{\mu} &=& 
x P^+ n_+^{\mu}
+p_T^{\mu}
+p^- n_-^{\mu} ,
\phantom{\frac{M}{P}} \\
S^{\mu} &=& 
S_L \, \frac{P^+}{M}n_+^{\mu}
+ S_T^{\mu} 
-S_L\, \frac{M}{2P^+}n_-^{\mu} ,
\label{e:vec}\\
T^{\mu \nu} &=& 
\frac{1}{2}\Biggl[ 
\frac{4}{3}\,S_{LL}\, \frac{{(P^+)}^2}{M^2}\, n_+^{\mu}n_+^{\nu}
+ \frac{P^+}{M}\,n_+^{\{ \mu} S_{LT}^{\nu \}} 
\nn \\ && \qquad \mbox{}
-\frac{2}{3}\,S_{LL}\,\left(n_+^{\{\mu} n_-^{\nu\}}-g_{T}^{\mu \nu} \right) 
+{S_{TT}^{\mu \nu}} 
\nn \\ && \qquad \mbox{}
- \frac{M}{2P^+} \,n_-^{\{ \mu} {S_{LT}^{\nu \}}}
+\frac{1}{3}\,S_{LL}\,\frac{M^2}{{(P^+)}^2}\, n_-^{\mu} n_-^{\nu} 
\Biggr]  \label{e:tens} 
\far

When only one hadron is considered, e.g.~in inclusive DIS, there is an
arbitrariness in the choice of $n_-$, though this does not affect physical
observables. In processes where another hadron is present, such as
one-particle inclusive leptoproduction, $n_-$ can be conveniently connected
 to the momentum of the produced hadron, so that
$P^{\mu} = P_h^- n_-^{\mu} + (M_h^2/2P_h^-)n_+^{\mu}$  .
This choice of light-like directions is particularly useful to analyze current
fragmentation in leptoproduction.  In this case one finds that up to 
order in $1/Q^2$ only {\em one} light-like component of  the hadron momentum is
relevant.   If we choose the relevant component of the target momentum to be
$P^+$,  then the relevant component of the outgoing hadron momentum will be
$P_h^-$.
We need to define the decomposition of the fragmenting quark momentum 
$k^{\mu} = (1/z) P_h^- n_-^{\mu}+k_T^{\mu} +k^+ n_-^{\mu}$, while
to obtain the decomposition for the outgoing hadron's spin vector and tensor,
it is sufficient to interchange the $+$ and $-$ components in
Eq.~(\ref{e:vec}) and Eq.~(\ref{e:tens}).

In semi-inclusive DIS one needs to consider the following integrated
correlation function:
\ar
\Phi(x ,{\bm p}_T)& =& \frac{1}{2} \left. \int \de p^-\;\Phi(p,P,S,T)
		\right|_{p^+= x P^+}\, , 
\label{e:dist}\\
\Delta(z ,{\bm k}_{T})& =& 
		\frac{1}{4z} \left. \int \de k^+ \;
                    \Delta(k,P_h,S_h,T_h) \right|_{k^-=\frac{P_h^-}{z}}\, .
\label{e:frag}
\far
In inclusive processes or after integrating the semi-inclusive cross sections
over the outgoing hadron's perpendicular momentum one needs to
consider the following ones:  
\ar
\Phi(x)& =& \frac{1}{2} \left. \int \de^2 {\bm p}_T\, \de p^-\;\Phi
	(k,P_h,S_h,T_h)	\right|_{p^+= x P^+} \, ,
\label{e:dist2}\\
\Delta(z)& =& 
		\frac{z}{4} \left. \int \de^2 {\bm k}_T \, \de k^+ \;
                    \Delta(k,P_h,S_h,T_h) \right|_{k^-=\frac{P_h^-}{z}}. 
\label{e:frag2}
\far
Note that in the case of fragmentation, it is conventional to integrate over
$-z\,\bm k_T$, which is the transverse momentum of the produced hadron with
respect to the quark. This can be checked by applying a Lorentz transformation
 that does not affect the minus component nor the integration over the plus 
component. Using coordinates $[a^-,a^+,\bm a_T]$, the required transformation
is 
\ar
\left[ P_h^-, \frac{M_h^2}{2P_h^-}, \bm 0_T \right]
&\quad \rightarrow \quad &
\left[ P_h^-, \frac{M^2 + z^2\,\bm k_T^2}{2P_h^-}, -z\,\bm k_T \right]
\\
\left[ \frac{P_h^-}{z}, k^+, \bm k_T \right]
&\quad \rightarrow \quad &
\left[ \frac{P_h^-}{z}, k^+  - \frac{z\,\bm k_T^2}{2P_h^-}, \bm 0_T \right] .
\far

We are going to separate different parts of the correlation functions
depending on the polarization conditions they require to be observed.
We will use the 
subscript $U$ to denote unpolarized hadrons, the subscript
$L$ and $T$ to denote respectively longitudinal and transverse vector
polarization and finally the subscripts $LL$, $LT$ and $TT$ to denote
longitudinal-longitudinal, longitudinal-transverse and transverse-transverse
tensor polarization.

In leading order in $1/Q$, the parameterization of the ${\bm p}_T$ 
dependent correlation function,
defined in Eq.~(\ref{e:dist}), is (we remind the reader that terms in round
brackets are T-odd)
\begin{eqnarray} 
\Phi_U (x, {\bm p}_T)& =& \frac{1}{4}\left\{
	f_1(x, p_T^2)\,\nslash_+ +
  \left(h_1^{\perp}(x, p_T^2)\,
		\sig_{\mu \nu}\frac{p_T^{\mu}}{M}  n_+^{\nu}\right) \right\},\\
\Phi_L (x, {\bm p}_T)& =& \frac{1}{4}\left\{
	g_{1L}(x, p_T^2)\,S_L \,\g_5 \nslash_+ +
	h_{1L}^{\perp}(x, p_T^2)\,S_L\,
		\ii \sig_{\mu \nu}\g_5 n_+^{\mu}\frac{ p_T^{\nu}}{M}\right\},\\
\Phi_T (x, {\bm p}_T)& =& \frac{1}{4}\left\{
	g_{1T}(x, p_T^2)\,\frac{{\bm S}_T \cdot {\bm p}_T}{M}\,\g_5 \nslash_+ +
	h_{1T}(x, p_T^2)\,
		\ii \sig_{\mu \nu} \g_5 n_+^{\mu} S_T^{\nu} \right. \nn \\
&&\left.\mbox{}+
	h_{1T}^{\perp}(x, p_T^2)\,\frac{{\bm S}_T \cdot {\bm p}_T}{M}\, 
	     \ii \sig_{\mu \nu}\g_5 n_+^{\mu}\frac{p_T^{\nu}}{M} \right. \nn \\
&&\left.\mbox{}+
  \left(f^{\perp}_{1T}(x, p_T^2)\,\eps_{\mu \nu \rho \sig}
	   \g^{\mu}n_+^{\nu}\frac{p_T^{\rho}}{M}S_T^{\sig} \right)\right\}, \\
\Phi_{LL} (x, {\bm p}_T)& =& \frac{1}{4}\left\{
	f_{1LL}(x, p_T^2)\,S_{LL} \,\nslash_+ +
  \left(h_{1LL}^{\perp}(x, p_T^2)\,S_{LL} \,
		\sig_{\mu \nu} \frac{p_T^{\mu}}{M}n_+^{\nu}\right) \right\}, \\
\Phi_{LT} (x, {\bm p}_T)& =& \frac{1}{4}\left\{
	f_{1LT}(x, p_T^2)\,\frac{{\bm S}_{LT} \cdot {\bm p}_T}{M}\, \nslash_+ +
  \left(g_{1LT}(x, p_T^2)\,\eps_T^{\mu \nu}S_{LT\,\mu}\frac{p_{T\,\nu}}{M} 
		\,\g_5 \nslash_+\right) 	\right. \nn \\ 
&&\left.\mbox{}+
  \left(h'_{1LT}(x, p_T^2)
    \, \ii \sig_{\mu \nu} \g_5 n_+^{\mu}\eps_T^{\nu \rho}S_{LT\,\rho}\right)
			\right. \nn \\ 
&&\left.\mbox{}+
  \left(\,h_{1LT}^{\perp}(x, p_T^2)\, \frac{{\bm S}_{LT} \cdot {\bm p}_T}{M}\,
			\sig_{\mu \nu}\frac{p_T^{\mu}}{M} n_+^{\nu} 
			\right) \right\},\\
\Phi_{TT} (x, {\bm p}_T)& =& \frac{1}{4}\left\{
   f_{1TT}(x, p_T^2)\,\frac{{\bm p}_T \cdot {\bm S}_{TT} \cdot {\bm p}_T}{M^2}
		\, \nslash_+  \right. \nn \\ 
&&\left.\mbox{}-
  \left(g_{1TT}(x, p_T^2)\, \eps_T^{\mu \nu} S_{TT\,\nu \rho}
	\frac{p_T^{\rho} p_{T\,\mu}}{M^2}\,\g_5 \nslash_+\right)\right. \nn \\ 
&&\left.\mbox{}-
  \left(h'_{1TT}(x, p_T^2)\, \ii \sig_{\mu \nu} \g_5 n_+^{\mu}
			\eps_T^{\nu \rho} S_{TT\,\rho \sig}
	\frac{p_T^{\sig}}{M} \right)
			\right. \nn \\ 
&&\left.\mbox{}+
  \left(\,h_{1TT}^{\perp}(x, p_T^2)\,
		\frac{{\bm p}_T \cdot {\bm S}_{TT} \cdot {\bm p}_T}{M^2}\,
		\sig_{\mu \nu} \frac{p_T^{\mu}}{M} n_+^{\nu} \right) \right\}.
\end{eqnarray} 

The parameterization of the correlation function after integration upon
${\bm p}_T$, as defined in Eq.~(\ref{e:dist2}), is
\begin{eqnarray} 
\Phi_U (x)& =& \frac{1}{4}\,
		f_1(x)\nslash_+, \\
\Phi_L (x)& =& \frac{1}{4}\,
		g_{1}(x)\,S_L\,\g_5 \nslash_+, \\
\Phi_T (x)& =& \frac{1}{4}\,
		h_{1}(x)\,
		\ii \sig_{\mu \nu} \g_5 n_+^{\mu} S_T^{\nu}, \\
\Phi_{LL} (x)& =& \frac{1}{4} f_{1LL}(x)\,S_{LL} \,\nslash_+, \\
\Phi_{LT} (x)& =& \frac{1}{4}\,\left(
		h_{1LT}(x)\,
    \ii \sig_{\mu \nu} \g_5 n_+^{\mu}\eps_T^{\nu \rho}S_{LT\,\rho}\right), 
			\label{e:ji2}\\
\Phi_{TT} (x)& =& 0, 
\end{eqnarray}
where
\begin{eqnarray}
g_1(x) &=& \int \de^2 {\bm p}_T \;g_{1L}(x,{p}^2_T),  \\
h_1(x) &=& \int \de^2 {\bm p}_T \;h_1(x,{p}^2_T) 
	\;=\;\int \de^2 {\bm p}_T \left (h_{1T} (x,{p}^2_T)
	+ \frac{{\bm p}_T^2}{2M^2}\;h_{1T}^{\perp}(x,{p}^2_T)\right), \\
h_{1LT}(x)
       &=& \int \de^2 {\bm p}_T \;h_{1LT}(x,{p}^2_T) 
	\;=\;\int \de^2 {\bm p}_T \left (h'_{1LT} (x,{p}^2_T)
	+ \frac{{\bm p}_T^2}{2M^2}\;h_{1LT}^{\perp}(x,{p}^2_T)\right). 
\end{eqnarray}

The decomposition of the correlation function
$\Delta$ is identical after the replacements $\{x,p_T,S,M,n_+\} \rightarrow  
\{z,k_T,S_h,M_h,n_-\}$ and the notation replacement 
$f \rightarrow D$, $g \rightarrow G$,  $h \rightarrow H$.

In App.~\ref{a:distr} all the possible distribution functions are  projected
 out of the complete correlation function. 
In Tab.~\ref{t:transv} and Tab.~\ref{t:notransv} we give a summary of all the
distribution functions, respectively before and after
integration upon ${\bm p}_T$.

The function $f_{1LL}$ has been already studied in \cite{hjm},
where it was given the name $b_1$ (Note that actually $f_{1LL}$ =
$-\frac{2}{3}\,b_{1}$). Although this name has been already used  also by
other authors (e.g. \cite{hk,sst}), we felt the need to change it to follow a
more systematic naming.  The function $f_{1LT}$ is
analogous to the function $c_1$ introduced in \cite{hk}, although the
different approach followed in that article requires a more careful comparison.

It is worthwhile to note that, as suggested by Eq.~(\ref{e:ji2}),
dealing with spin-1 particles offers the possibility of measuring a
time-reversal odd function in leading twist and without considering intrinsic
transverse momentum. The particular fragmentation function $H_{1LT}$,
equivalent to the distribution function $h_{1LT}$, has been introduced in
\cite{ji}, where it was named $\^h_{\bar{1}}$.

\renewcommand{\arraystretch}{1.5}
\begin{table}[p]\center
\begin{minipage}{11cm}
\begin{tabular}{|c|c|c|c|c|c|c|}
 	& \multicolumn{2}{c|}{$[\g^+]$}		
		&\multicolumn{2}{c|}{$[\g^+ \g_5]$}	&
			\multicolumn{2}{c|}
{$[\ii \sig^{i+} \g_5]$}	
\\
\hline
	&TR-even&TR-odd	&TR-even&TR-odd	&TR-even&TR-odd	\\
\hline
U	&$f_1$	&	& 	&	& &$(h_{1}^{\perp})$ \\
L	&	&	&$g_{1L}$&	& $h_{1L}^{\perp}$&	\\
T	& &$(f_{1T}^{\perp})$&$g_{1T}$&	& \begin{tabular}{cc}
						$h_{1T}$&$ h_{1T}^{\perp}$
						\end{tabular}&	\\
LL	&$f_{1LL}$&	&	&	& &$(h_{1LL}^{\perp})$ \\
LT	&$f_{1LT}$&	& &$(g_{1LT})$	& & \begin{tabular}{cc}
	  				$(h'_{1LT}$&$h_{1LT}^{\perp})$
	     				\end{tabular} \\
TT	&$f_{1TT}$&	& &$(g_{1TT})$	& &\begin{tabular}{cc}
	  				$(h'_{1TT}$&$h_{1TT}^{\perp})$
	     				\end{tabular} \\ 
\end{tabular}\mbox{}\\
\end{minipage}
\caption{List of leading twist distribution functions, divided in
	time-reversal even and time-reversal odd} 
\label{t:transv}
\end{table}
\begin{table}[p]
\center
\begin{minipage}{10cm}
\begin{tabular}{|c|c|c|c|c|c|c|}
 	& \multicolumn{2}{c|}{$[\g^+]$}		
		&\multicolumn{2}{c|}{$[\g^+ \g_5]$}	&
			\multicolumn{2}{c|}{$[\ii \sig^{i+} \g_5]$}	\\
\hline
	&TR-even&TR-odd	&TR-even&TR-odd	&TR-even&TR-odd	\\
\hline
U	&$f_1$	&	& 	&	&	&	\\
L	&	&	&$g_{1}$&	&	&	\\
T	&	&	&	&	& $h_{1}$&	\\
LL	&$f_{1LL}$&	&	&	&	&	\\
LT	&	&	&	&	&	&$(h_{1LT})$	\\
TT	&	&	&	&	&	&	 \\ 
\end{tabular}\mbox{}\\
\end{minipage}
\caption{List of remaining leading twist distribution functions after
	integration upon ${\bm p}_T$, divided in
	time-reversal even and time-reversal odd} 
\label{t:notransv}
\end{table}

It is sometimes useful (for instance for calculation of azimuthal asymmetries)
 to consider the $p_T^{\alpha}$-weighted function
\begin{equation}
\frac{1}{M} \Phi_{\partial}^{\alpha} (x) \equiv 
	\int \de^2{\bm p}_T \, \frac{p_T^{\alpha}}{M} \, \Phi(x,{\bm p}_T).
\end{equation}
Non vanishing at twist two we have
\begin{eqnarray}
\frac{1}{M}\left(\Phi_{\partial}^{\alpha}\right)_U (x)& =& 
-\frac{1}{4}
	\left(h_1^{\perp (1)}(x)\,
		\sig^{\alpha \nu}  n_{+\,\nu}\right),\\
\frac{1}{M}\left(\Phi_{\partial}^{\alpha}\right)_L (x)& =&  
-\frac{1}{4}\;
	h_{1L}^{\perp (1)}(x)\,S_L\,
		\ii \sig^{\mu \alpha}\g_5 n_{+\,\mu},\\
\frac{1}{M}\left(\Phi_{\partial}^{\alpha}\right)_T (x)& =&  
-\frac{1}{4}\left\{
	g_{1T}^{(1)}(x)\, S_T^{\alpha}\,\g_5 \nslash_+ +
  \left(f^{\perp (1)}_{1T}(x)\,\eps^{\mu \nu \alpha \sig}
	   \g_{\mu}n_{+\,\nu} S_{T\,\sig} \right)\right\}, \\
\frac{1}{M}\left(\Phi_{\partial}^{\alpha}\right)_{LL} (x)& =&  
-\frac{1}{4}
  \left(h_{1LL}^{\perp (1)}(x)\,S_{LL} \,
		\sig^{\alpha \nu} n_{+\,\nu}\right), \\
\frac{1}{M}\left(\Phi_{\partial}^{\alpha}\right)_{LT} (x)& =&  
-\frac{1}{4}\left\{
	f_{1LT}^{(1)}(x)\,S_{LT}^{\alpha}\, \nslash_+ +
  \left(g_{1LT}^{(1)}(x)\,\eps_T^{\mu \alpha}S_{LT\,\mu} 
		\,\g_5 \nslash_+\right) 	\right\} , \\
\frac{1}{M}\left(\Phi_{\partial}^{\alpha}\right)_{TT} (x)& =& 
\frac{1}{4}
	\left(h_{1TT}^{(1)}(x)\,S_{TT}^{\alpha \mu} 
		\,\sig_{\mu \nu} n_+^{\nu}\right),
\end{eqnarray}
where we used the notation
\begin{equation}
h_{1}^{\perp (1)}(x)
       = \int \de^2 {\bm p}_T \;h_{1}^{\perp (1)}(x,{p}^2_T) 
	= \int \de^2 {\bm p}_T\, \frac{{\bm p}_T^2}{2M^2}
				\;h_{1}^{\perp}(x,{p}^2_T), 
\end{equation}
and we introduced the function
\begin{equation}
h_{1TT}(x,{p}^2_T) 
	= h'_{1TT} (x,{p}^2_T)
	+ \frac{{\bm p}_T^2}{2M^2}\;h_{1TT}^{\perp}(x,{p}^2_T).
\end{equation}	
	  
\section{Semi-inclusive cross sections with tensor polarization in the final
state} 
\label{s:cross}

We consider one-particle inclusive DIS events where the
target consists of a spin-$\frac{1}{2}$ hadron and the fragment is a spin-1
hadron with tensor polarization only.  
We allow only time-reversal odd fragmentation functions to occur,
assuming  that there are no time-reversal odd distribution function.

A short note on the kinematics is the first necessary ingredient.
In Sec.~\ref{s:main} we defined with the help of the momenta $P$ and $P_h$
the transverse projector $g_T^{\mu \nu}$ and transverse vectors. From the
experimental point of view it is customary to work with vectors constructed
from the momenta $q$ and $P$. They are used
to define a space-like direction $\^q^\mu = q^\mu/Q$, an orthogonal time-like
direction $\^t^{\mu} = \frac{1}{Q} (2 {\xbj} P^{\mu} + q^{\mu})$, where
$\xbj=Q^2/2 (P \cdot q)$ (neglecting
mass corrections of order $1/Q^2$), and perpendicular directions via the tensor
\begin{equation}
g_{\perp}^{\mu \nu}= g^{\mu \nu} + \^q^{\mu} \^q^{\nu} - \^t^{\mu} \^t^{\nu}.
\label{e:gperp}
\end{equation}
After introducing the scaling variable $z_h = 2 P_h \cdot q / Q^2 \simeq 
P \cdot P_h / P \cdot q$  (neglecting order $1/Q^2$ corrections) 
and using $g_T^{\mu\nu}$ or 
$g_\perp^{\mu\nu}$ we can write the relation
\begin{equation}
\xbj P^\mu - \frac{P_h^\mu}{z_h} + q^\mu = q_T^\mu 
		= - \frac{P_{h\perp}^\mu}{z_h},
\label{covtransf}
\end{equation}
showing that the combination on the left-hand side 
is either the transverse component of $q$ (since $P_T = P_{hT}$ = 0) or the
perpendicular component of $-P_h/z_h$ (since $P_\perp = q_\perp$ = 0).
 
To explicitly write cross sections we also need  the scaling
variable 
$y={P \cdot q}/{P \cdot l}$,
where $l$ denotes the incoming lepton momentum, and the azimuthal
angle $\phi^\ell$ of the lepton scattering plane. Cross sections will be
differential with respect to the variables $\xbj$, $z_h$, $y$, $\phi^\ell$ and
${\bm P}_{h\perp}$.  When they do not vanish, we will also give cross sections
integrated over ${\bm P}_{h \perp}$ and $\phi^\ell$. 
The general formula is
\begin{equation}
\frac{2\pi\,\de \sig (l+H \rightarrow l'+h+X)}
	{\de \phi^\ell \de \xbj \de z_h \de y \de^2 {\bm P}_{h \perp}}=
	\frac{\pi \alpha^2}{2Q^4}\;\frac{y}{z_h}\; L_{\mu \nu}
	\; 2M W^{\mu \nu}, 
\end{equation}
where $L_{\mu \nu}$ is the lepton tensor and $W^{\mu \nu}$ is
the hadronic tensor given by the convolution of the soft parts,
\begin{eqnarray}
2 M W^{\mu \nu}&=&2z_h \int \de^2 {\bm p}_T \de^2 {\bm k}_T 
	 \ \delta^2 ({\bm p}_T + {\bm q}_T - {\bm k}_T) 
	\;\tr \left[2 \Phi(\xbj,{\bm p}_T)\, \g^{\mu} 
		\; 2 \Delta(z_h,{\bm k}_T)\, \g^{\nu} \right],
\label{tensor}
\end{eqnarray}
where it is understood that a summation over the charge squared weighted sum 
over quark flavors has to be included. 
The full form of the hadronic tensor can be obtained by introducing the
correlation functions described in the previous section (see
App.~\ref{a:hadronic}).
To shorten the formulae we will use the notation
\begin{equation}
{\bm I}\left[ \ldots \right] = 2z_h \int \de^2 {\bm p}_T \de^2 {\bm k}_T \,
	 \delta^2 ({\bm p}_T + {\bm q}_T - {\bm k}_T) \ldots.
\end{equation}

It is convenient to express the perpendicular vectors with respect 
to the only measured perpendicular direction, i.e. that of 
${\bm P}_{h \perp}$, the outgoing hadron's perpendicular momentum. 
Defining the unit vector in this direction 
$\^{\bm h}=\frac{{\bm P}_{h \perp}}{|P_{h \perp}|}$,
we are then going to use the following notation
\begin{eqnarray}
a_{\perp}^{\mu} 
	&=& a_x \,\^h^{\mu}
	   + a_y \,\eps_{\perp}^{\mu \nu}\^h_{\nu}.
\end{eqnarray}

As it has been
shown in \cite{multa}, the difference between $g_T^{\mu \nu}$ in
Eq.~(\ref{e:gtrans}) and $g_{\perp}^{\mu \nu}$ in Eq.~(\ref{e:gperp})
is of order 1/Q, i.e. (neglecting order $1/Q^2$ parts) 
\begin{equation}
g_\perp^{\mu\nu} = g_T^{\mu\nu} - \frac{\sqrt{2}n_+^{\{\mu}q_T^{\nu\}}}{Q}
\qquad \mbox{or} \qquad
g_T^{\mu\nu} = g_{\perp}^{\mu\nu}-\frac{Q_T}{Q}\sqrt{2}n_+^{\{\mu} \^h^{\nu\}},
\label{e:gmunu}
\end{equation}
where $Q_T = |P_{h\perp}|/z_h$.
This relation 
implies that if we already have projected out a transverse vector, 
the additional projection with $g_\perp^{\mu\nu}$ does not give additional 
terms, i.e. $a_{T\perp}^\mu$ = $g_\perp^{\mu\rho}a_{T\rho} = a_T^\mu$, even if
$a_{\perp}^\mu  \not= a_T^\mu$ (see App.~\ref{a:decay}). 
This is true up to corrections of order $1/Q^2$.

We will indicate as $\phi_S^h$ the angle between ${\bm S}_T$
 and ${\bm P}_{h \perp}$,
as $\phi_h^\ell$ the angle between ${\bm P}_{h \perp}$ 
and the scattering plane, 
 as $\phi_S^\ell$ the angle between ${\bm S}_T$ and the 
scattering plane.

For the tensor $T_h$ 
we introduce azimuthal angles defined as:
\begin{eqnarray} 
\tan{(\phi_{h\,LT}^h)}&=&\;\tan{(\phi_{h\,LT}^\ell-\phi_h^\ell)}\;
 =\; \frac{S_{h\,LT}^{y}}{S_{h\,LT}^{x}}\;, \nn \\
\tan{(2 \phi_{h\,TT}^h)}&=&\tan{(2 \phi_{h\,TT}^\ell- 2 \phi_h^\ell)} \;
=\; \frac {S_{h\,TT}^{xy}}{S_{h\,TT}^{xx}}\;,
\end{eqnarray} 
and the the quantities
\begin{equation}
|S_{h\,LT}|\;=\; \sqrt{(S_{h\,LT}^{x})^2 +  (S_{h\,LT}^{y})^2}\;, \qquad
|S_{h\,TT}|\;=\; \sqrt{(S_{h\,TT}^{xx})^2 +  (S_{h\,TT}^{xy})^2}\;.
\end{equation}

In a real experiment, where polarimetry is performed on the final-state hadron,
cross section will not depend on the spin tensor parameters but rather on the
analyzing powers determined from the momenta of decay products. We omit
writing explicit differential cross sections in terms of the momenta of the
decay products, but we merely point out that spin tensor parameters in cross
section formulae must be replaced by the corresponding analyzing powers, as
given in App.~\ref{a:decay}.

\subsection{Unpolarized lepton beam and unpolarized target ($UU$)}

In this case, the differential cross section is
\begin{eqnarray}
\lefteqn{\frac{\de \sig_{{\scriptscriptstyle UU}}
	(l+H \rightarrow l'+\vec{\vec{h}}+X)}{\de \xbj \de z_h
		\de y \de^2 {\bm P}_{h \perp}}=}
\nn \\&&
	\frac{4 \pi \alpha^2 s}{Q^4}\;\left(1-y-\frac{y^2}{2}\right) \xbj
	\;\Biggl\{ S_{h\,LL} \;{\bm I} \left[f_1 \; D_{1LL}\right]  
\nn \\&&  \;\; \mbox{}+
 |S_{h\,LT}|\,\cos{(\phi_{h\,LT}^h)} 
	\;{\bm I} \left[\frac{ k^x}{M_h}\, f_1 \; D_{1LT}\right]
\nn \\&&  \;\; \mbox{}\left.+
	|S_{h\,TT}|\,\cos{(2\phi_{h\,TT}^h)}\;{\bm I} \left[
		\frac{(k^x)^2-(k^y)^2}{M_h^2}\; f_1 \; D_{1TT}\right]\right\},
\end{eqnarray}
while after integration over ${\bm P}_{h\perp}$ the differential cross 
section is
\begin{equation}
\frac{\de \sig_{{\scriptscriptstyle UU}}
	(l+H \rightarrow l'+\vec{\vec{h}}+X)}{\de \xbj \de z_h \de y}=
 \frac{4 \pi \alpha^2 s}{Q^4}\;\left(1-y-\frac{y^2}{2}\right) \xbj\; 
		S_{h\,LL} \; f_1(\xbj)  \; D_{1LL}(z_h).
\end{equation}

\subsection{Polarized lepton beam and unpolarized target ($LU$)}

Indicating with $\lambda_e$ the helicity of the incoming lepton, the 
differential cross section is
\begin{eqnarray}
\lefteqn{\frac{\de \sig_{{\scriptscriptstyle LU}}
	(\vec{l}+H \rightarrow l'+\vec{\vec{h}}+X)}{\de \xbj \de z_h
		\de y \de^2 {\bm P}_{h \perp}}=}
\nn \\&&
	\frac{4 \pi \alpha^2 s}{Q^4}\;\lambda_e\,
		 y\left(1-\frac{y}{2}\right) \xbj \;\Biggl\{
	|S_{h\,LT}|\,\sin{(\phi_{h\,LT}^h)}
		\;{\bm I} \left[\frac{k^x}{M_h}\; f_1 \; G_{1LT}\right] 
\nn \\&&\;\;\mbox{}+
	|S_{h\,TT}|\,\sin{(2\phi_{h\,TT}^h)}\;{\bm I} 
       \left[\frac{(k^x)^2-(k^y)^2}{M_h^2}\; f_1 \; G_{1TT} \right]\Biggr\}.
\end{eqnarray}

\subsection{Unpolarized lepton beam and longitudinally polarized target ($UL$)}

\begin{eqnarray}
\lefteqn{\frac{2\pi\,\de \sig_{{\scriptscriptstyle UL}}
	(l+\vec{H} \rightarrow l'+\vec{\vec{h}}+X)}{\de \phi^\ell \de \xbj 
		\de z_h \de y \de^2 {\bm P}_{h \perp}}=}
\nn \\&&
	\frac{4 \pi \alpha^2 s}{Q^4}\;\xbj\, 
			\left(1-y-\frac{y^2}{2}\right) S_L 
	\Biggl\{|S_{h\,LT}|\,\sin{(\phi_S^\ell -\phi_h^\ell)}
		\;{\bm I} \left[\frac{k^x}{M_h}\; g_{1L} \; G_{1LT}\right]
\nn \\&& \mbox{} \;\; +
	|S_{h\,TT}|\,\cos{(2\phi_{h\,TT}^h)}\;{\bm I} 
     \left[\frac{(k^x)^2-(k^y)^2}{M_h^2}\; g_{1L}\; G_{1TT} \right]\Biggr\}
							 \nn \\
&&     \mbox{}+\frac{4 \pi \alpha^2 s}{Q^4}\;\xbj\, (1-y)\; S_L\; 
	\Biggl\{
	|S_{h\,LT}|\,\sin{(\phi_{h\,LT}^\ell+\phi_h^\ell)}
	\;{\bm I} \left[\frac{p^x}{M} \; h_{1L}^{\perp} \,H_{1LT}\right]
      \nn \\&& \;\;\mbox{}+ 
	|S_{h\,TT}|\,\sin{(2\phi_{h\,TT}^\ell)}
	\;{\bm I} \left[\frac{{\bm p} \cdot {\bm k}}{M M_h} 
			\, h_{1L}^{\perp} \, H_{1TT}\right]+ 
	S_{h\,LL}\,\sin{(2\phi_h^\ell)}
	\;{\bm I} \left[\frac{ p^x k^x -p^y k^y}{M M_h} 
				\; h_{1L}^{\perp} \,H_{1LL}^{\perp}\right]
      \nn \\&& \;\;\mbox{}- 
	|S_{h\,LT}|\,\sin{(\phi_{h\,LT}^\ell-3\phi_h^\ell)}
	\;{\bm I}\left[ \frac{p^x\left[(k^x)^2 -(k^y)^2\right] 
			-2\,k^x k^y p^y }{2 M M_h^2} 
				\; h_{1L}^{\perp} \,H_{1LT}^{\perp}\right] 
      \nn \\&& \;\;\mbox{}- 
	|S_{h\,TT}|\,\sin{(2\phi_{h\,TT}^\ell-4\phi_h^\ell)}
	\;{\bm I}\left[\frac{ 2\left[(k^x)^2 -(k^y)^2)\right]
	   (k^x p^x - k^y p^y)-{\bm k_T}^2 ({\bm p} \cdot {\bm k})}{2M M_h^3} 
			\; h_{1L}^{\perp} \,H_{1TT}^{\perp}\right] \Biggr\}.
\end{eqnarray}

\subsection{Polarized lepton beam and longitudinally polarized target ($LL$)}

\begin{eqnarray}
\lefteqn{\frac{\de \sig_{{\scriptscriptstyle LL}}
	(\vec{l}+\vec{H} \rightarrow l'+\vec{\vec{h}}+X)}{\de \xbj \de z_h
		\de y \de^2 {\bm P}_{h \perp}}=}
\nn \\&&
	\frac{4 \pi \alpha^2 s}{Q^4}\; 2\, 
	\lambda_e \, S_L \,\xbj \,y \left(1-\frac{y}{2}\right)
	\;\Biggl\{ S_{h\,LL} \;{\bm I} \left[g_{1L} \; D_{1LL}\right] 
\nn \\&&  \;\; \mbox{} +
	|S_{h\,LT}|\,\cos{(\phi_{h\,LT}^h)} 
		\;{\bm I} \left[\frac{ k^x}{M_h}\, g_{1L} \; D_{1LT}\right] 
\nn \\&&  \;\; \mbox{} +|S_{h\,TT}|\,\cos{(2\phi_{h\,TT}^h)}
	\;{\bm I} \left[
	      \frac{(k^x)^2-(k^y)^2}{M_h^2}\; g_{1L} \; D_{1TT}\right]\Biggr\}.
\end{eqnarray}

\subsection{Unpolarized lepton beam and transversely polarized target ($UT$)}

\begin{eqnarray}
\lefteqn{\frac{2\pi\,\de \sig_{{\scriptscriptstyle LT}}
	(l+\vec{H} \rightarrow l'+\vec{\vec{h}}+X)}{\de \phi^\ell \de \xbj 
		\de z_h \de y \de^2 {\bm P}_{h \perp}}=}
\nn \\&&
	\frac{4 \pi \alpha^2 s}{Q^4}\;
		\xbj \left(1-y-\frac{y^2}{2}\right) |S_{T}|\;
	\Biggr\{|S_{h\,LT}|\,\cos{(\phi_S^\ell-\phi_h^\ell)}
	\,\sin{(\phi_{h\,LT}^\ell-\phi_h^\ell)}
     \;{\bm I} \left[\frac{p^x k^x }{M M_h}\; g_{1T} \; G_{1LT}\right] 
\nn \\&&	\;\; \mbox{}
	+|S_{h\,TT}|\,\cos{(\phi_S^\ell-\phi_h^\ell)}
	\,\sin{(\phi_{h\,TT}^\ell-\phi_h^\ell)}
	   \;{\bm I} \left[\frac{p^x \left[(k^x)^2-(k^y)^2\right] }
			{M M_h^2}\; g_{1T} \; G_{1TT}
					\right] 
\nn \\ &&\;\;\mbox{}+|S_{h\,LT}|\,\sin{(\phi_S^\ell-\phi_h^\ell)}
	\,\cos{(\phi_{h\,LT}^\ell-\phi_h^\ell)}
     \;{\bm I} \left[\frac{p^y k^y }{M M_h}\; g_{1T} \; G_{1LT}\right]
\nn \\&&	\;\; \mbox{}
	+|S_{h\,TT}|\,\sin{(\phi_S^\ell-\phi_h^\ell)}
	\,\cos{(\phi_{h\,TT}^\ell-\phi_h^\ell)}
	   \;{\bm I} \left[\frac{2\,p^x k^x k^y}{M M_h^2}\; g_{1T} \; G_{1TT}
					\right] \Biggr\}\nn \\
&&	\mbox{}+\frac{4 \pi \alpha^2 s}{Q^4}\;
	    \xbj\;(1-y) \;|S_{T}|\;
	\Biggr\{|S_{h\,LT}|\;\sin{(\phi_{h\,LT}^\ell+\phi_{h\,T}^\ell)}
		\;{\bm I} \left[h_{1}\; H_{1LT}\right] 
\nn \\ && \;\;\mbox{}+|S_{h\,TT}|\;
	\sin{(2\phi_{h\,TT}^\ell+\phi_S^\ell-\phi_h^\ell)}
		\;{\bm I} \left[\frac{k^x}{M_h}\; h_{1}\; H_{1TT}\right]
\nn \\ && \;\;\mbox{}+ S_{h\,LL}\;\sin{(\phi_S^\ell+\phi_h^\ell)}
		\;\;{\bm I}\left[\frac{k^x}{M_h}\;h_{1}\;H_{1LL}^{\perp}\right]
\nn \\ && \;\;\mbox{}-|S_{h\,LT}|\;
	\sin{(\phi_{h\,LT}^\ell-\phi_S^\ell-2\phi_h^\ell)}
		\;{\bm I} \left[\frac{(k^x)^2 -(k^y)^2}{2M_h^2}
			\; h_{1} \; H_{1LT}^{\perp}\right] 
\nn \\ && \;\;\mbox{}+|S_{h\,LT}|\;
	\sin{(\phi_{h\,LT}^\ell-\phi_S^\ell+2\phi_h^\ell)}
		\;{\bm I} \left[\frac{(p^x)^2 -(p^y)^2}{2M^2}
			\; h_{1T}^{\perp} \; H_{1LT}\right]
\nn \\ && \;\;\mbox{}-|S_{h\,TT}|\;
	\sin{(2\phi_{h\,TT}^\ell-\phi_S^\ell-3\phi_h^\ell)}
		\;{\bm I} \left[\frac{k^x\, 
		\left[(k^x)^2 - (k^y)^2 -\frac{{\bm k}_T^2}{2}\right]}{M_h^3}
			\;h_{1} \; H_{1TT}^{\perp}\right]
\nn \\ && \;\;\mbox{}+|S_{h\,TT}|\;
	\sin{(2\phi_{h\,TT}^\ell-\phi_S^\ell+\phi_h^\ell)}
		\;{\bm I} \left[\frac{k^x\, \left[(p^x)^2 -(p^y)^2\right] 
				+ 2\, p^x p^y k^y}{2 M^2 M_h}
			\;h_{1T}^{\perp} \; H_{1TT}\right]
\nn \\ && \;\;\mbox{}-S_{h\,LL}\;\sin{(\phi_S^\ell-3\phi_h^\ell)}
		\;\;{\bm I}\left[\frac{k^x\, \left[(p^x)^2 -(p^y)^2\right] 
				- 2\, p^x p^y k^y}{2 M^2 M_h}
			\;h_{1T}^{\perp} \; H_{1LL}^{\perp}\right]
\nn \\ && \;\;\mbox{}-|S_{h\,LT}|\;
	\sin{(\phi_{h\,LT}^\ell+\phi_S^\ell-4\phi_h^\ell)}
		\;{\bm I} \left[\frac{\left[(k^x)^2 -(k^y)^2\right]
	\left[(p^x)^2 -(p^y)^2\right]-4\,p^x p^y k^x k^y}{4 M^2  M_h^2}\;
			h_{1T}^{\perp} \; H_{1LT}^{\perp}\right]
\nn \\ && \;\;\mbox{}-|S_{h\,TT}|\;
	\sin{(2\phi_{h\,TT}^\ell-\phi_S^\ell-3\phi_h^\ell)}
\nn \\ &&\;\;\left.\mbox{}\times 
	{\bm I} \left[\frac{k^x\left[(p^x)^2 -(p^y)^2\right]
	\left[(k^x)^2 - (k^y)^2-\frac{{\bm k}_T^2}{2}\right]
	-2\,p^x p^y k^y\,\left[(k^x)^2 - (k^y)^2 +\frac{{\bm k}_T^2}{2}\right]}
			{2 M^2  M_h^3}\;
			h_{1T}^{\perp} \; H_{1TT}^{\perp}\right]
				\right\}.\nn \\
\end{eqnarray}

After performing the integration over ${\bm P}_{h \perp}$ we obtain the cross
section:
\begin{eqnarray}
\lefteqn{\frac{2 \pi\, \de \sig_{{\scriptscriptstyle LT}}
	(l+\vec{H} \rightarrow l'+\vec{\vec{h}}+X)}
		{\de \phi^\ell \de \xbj \de z_h\de y}=} \nn \\
&&  \qquad \frac{4 \pi \alpha^2 s}{Q^4}\;\xbj\;(1-y)\;|S_{T}|\, |S_{h\,LT}|
	\sin{(\phi_{h\,LT}^\ell+\phi_S^\ell)}
		\; h_{1}(\xbj)  \; H_{1LT}(z_h).
\end{eqnarray}

We want to point out the importance of this last case, which would allow the
measurement of the chiral odd distribution function $h_1$ together
with a time-reversal odd and chiral odd fragmentation function,
requiring neither contributions non-leading in 1/Q, nor the
measurement of the transverse momentum of the outgoing hadron.

\subsection{Polarized lepton beam and transversely polarized target ($LT$)}

\begin{eqnarray}
\lefteqn{\frac{2\pi\,\de \sig_{{\scriptscriptstyle LT}}
	(\vec{l}+\vec{H} \rightarrow l'+\vec{\vec{h}}+X)}{\de \phi^\ell 
		\de \xbj \de z_h \de y \de^2 {\bm P}_{h \perp}}=}
\nn \\&&
	\frac{4 \pi \alpha^2 s}{Q^4}\; 2 \,\lambda_e \;\xbj\; 
	y \;\left(1-\frac{y}{2}\right)|S_{T}|
	\;\Biggl\{ S_{h\,LL} \,\cos{(\phi_S^\ell-\phi_h^\ell)}\,
		\;{\bm I} \left[\frac{p^x}{M}g_{1T} \; D_{1LL}\right]
\nn \\&& \; \;\mbox{}+ |S_{h\,LT}|\,
	\cos{(\phi_S^h)}\,\cos{(\phi_{h\,LT}^\ell-\phi_h^\ell)}
		\;{\bm I}\left[\frac{p^x k^x}{M M_h}\, g_{1T} \;D_{1LT}\right]
\nn \\ && \;\;\mbox{} 
		+|S_{h\,TT}|\,\cos{(\phi_S^h)}\,
		\cos{(2\phi_{h\,TT}^h)}
		\;{\bm I} \left[\frac{p^x\,[(k^x)^2-(k^y)^2]}{M M_h^2}
				\; g_{1T} \; D_{1TT}\right]
\nn \\&& \; \; \mbox{}+
	|S_{h\,LT}|\,\sin{(\phi_S^h)}\,
	\sin{(\phi_{h\,LT}^h)} 
	\;{\bm I} \left[\frac{p^y k^y}{M M_h}\, g_{1T} \; D_{1LT}\right] 
\nn \\&& \;\;\mbox{}+|S_{h\,TT}|\,\sin{(\phi_S^h)}\,
	\sin{(2\phi_{h\,TT}^h)}
	\;{\bm I} \left[\frac{2\,p^y k^x k^y}{M M_h^2}
				\; g_{1T} \; D_{1TT}\right]\Biggr\}
\end{eqnarray}

\section{Conclusions}

In this paper we have studied quark distribution and fragmentation functions
for hadrons with spin one. We have given a complete list of the functions that
can appear at leading order in $1/Q$ in electroweak hard processes. We have
included intrinsic transverse momentum dependence, useful for the treatment
of processes
in which more than one hadron is involved, such as 1-particle inclusive
leptoproduction. We have included time-reversal odd functions.  
In particular, 
time-reversal odd fragmentation
functions show up in single spin asymmetries. We have not estimated the
various functions, since  they contain  soft physics and as such are 
uncalculable at present. At best some positivity bounds can be given and 
issues like scale
dependence may be studied. Some of these aspects will be addressed in future
studies.

Our treatment is complete, allowing the calculation of inclusive
and semi-inclusive leptoproduction involving spin one hadrons in initial or
final state at tree-level and up to leading order in $1/Q$, but including the
full spin structure in initial (beam and target polarization) or final state
(polarimetry).   

In Sec.~\ref{s:cross} we have focussed on the specific process of
deep-inelastic leptoproduction of vector mesons ($\rho$ mesons) for which
polarimetry is possible from the analysis of the decay products ($\pi\pi$
final state). We calculated all cross-sections measurable with different beam
and target polarization. 

Amongst the results, we want to emphasize that vector meson leptoproduction
off transversely polarized nucleons allows the observation of the chiral-odd
transverse-spin distribution, $h_1(x)$ in a single spin asymmetry
involving the time-reversal odd fragmentation function, $H_{1LT}(z)$. Unlike
the situation involving spin 1/2 particles, this does not require any
azimuthal asymmetries, although the function $H_{1LT}$ itself is not known.

\acknowledgements{We would like to thank Daniel Boer and Elliot Leader for
fruitful discussions.
This work is supported by the Foundation for Fundamental
Research on Matter (FOM) and the Dutch Organization for Scientific Research 
(NWO).}

\appendix
\section{Interpretation of the components of the spin tensor}
\label{a:tensor}

A particular component of the spin tensor measures a combination of
probabilities of finding the system in a certain spin state (defined in the
particle rest frame). 

As ``analyzing'' spin states we can choose the eigenstates of the spin vector
operator in a particular direction. 
We can write the spin vector operator in terms of polar and azimuthal angles,
\begin{equation}
{\bm \Sigma}^i \^n_i = {\bm \Sigma}_x \cos{\theta}\cos{\varphi}
			+{\bm \Sigma}_y \cos{\theta}\sin{\varphi} 
			+{\bm \Sigma}_z \sin{\theta},
\end{equation}
and we can denote its eigenstates as $|m_{(\theta,\varphi)}\>$, $m$
being their magnetic quantum number.
The probability of finding one of these states can be calculated as
\begin{equation}
P\left(m_{(\theta,\varphi)}\right)= 
	\tr \left\{{\bm \rho} \, 
		|m_{(\theta,\varphi)}\>\<m_{(\theta,\varphi)}|\right\}.
\end{equation}

Inserting in Eq.~(\ref{e:density}) the spin tensor,
Eq.~(\ref{e:tensor}), and the spin vector, Eq.~(\ref{e:vector}),
the explicit form of the spin density matrix $\rho$
turns out to be
\begin{equation}
{\bm \rho}=\left(\begin{array}{ccc}
\frac{1}{3}+\frac{S_{LL}}{3}+\frac{S_L}{2} &
	 \frac{S_{LT}^{x} - \ii S_{LT}^{y}}{2\sqrt{2}}
			+\frac{S_{T}^{x} - \ii S_{T}^{y}}{2\sqrt{2}} &
				\frac{S_{TT}^{xx} - \ii S_{TT}^{xy}}{2}\\
\frac{S_{LT}^{x} + \ii S_{LT}^{y}}{2\sqrt{2}}
	+\frac{S_{T}^{x} + \ii S_{T}^{y}}{2\sqrt{2}} &
		\frac{1}{3}-\frac{2 S_{LL}}{3} &
			\frac{-S_{LT}^{x} + \ii S_{LT}^{y}}{2\sqrt{2}}
				+\frac{S_{T}^{x} - \ii S_{T}^{y}}{2\sqrt{2}}\\
\frac{S_{TT}^{xx} + \ii S_{TT}^{xy}}{2} &
	\frac{-S_{LT}^{x} - \ii S_{LT}^{y}}{2\sqrt{2}}
			+\frac{S_{T}^{x} + \ii S_{T}^{y}}{2\sqrt{2}}&
				\frac{1}{3}+\frac{S_{LL}}{3}-\frac{S_L}{2}
\end{array}\right). 
\label{e:rho}
\end{equation}
{}From this explicit formula one can check that
\begin{eqnarray}
S_{LL} &=& \frac{1}{2}P\left(1_{(0,0)}\right)
	 +\frac{1}{2}P\left(-1_{(0,0)}\right) - P\left(0_{(0,0)}\right), \nn \\
S_{LT}^{x} &=& P\left(0_{(-\frac{\pi}{4},0)}\right) 
		- P\left(0_{(\frac{\pi}{4},0)}\right), \nn \\
S_{LT}^{y} &=& P\left(0_{(-\frac{\pi}{4},\frac{\pi}{2})}\right)
		- P\left(0_{(\frac{\pi}{4},\frac{\pi}{2})}\right), \nn \\
S_{TT}^{xx} &=& P\left(0_{(\frac{\pi}{2},-\frac{\pi}{4})}\right) 
		- P\left(0_{(\frac{\pi}{2},\frac{\pi}{4})}\right), \nn \\
S_{TT}^{xy} &=& P\left(0_{(\frac{\pi}{2},\frac{\pi}{2})}\right) 
		- P\left(0_{(\frac{\pi}{2},0)}\right). \nn 
\end{eqnarray}

Below, we suggest a diagrammatic interpretation of these probability
combinations.  
Arrows represent spin states
$m=+1$ and $m=-1$ in the direction of the arrow itself, while dashed lines
denote spin state $m=0$ again in the direction of the line itself.
\begin{eqnarray*}
	& \makebox[1cm][c]{\makebox[4cm][r]{\parbox[c]{3cm}
		{\epsfig{figure=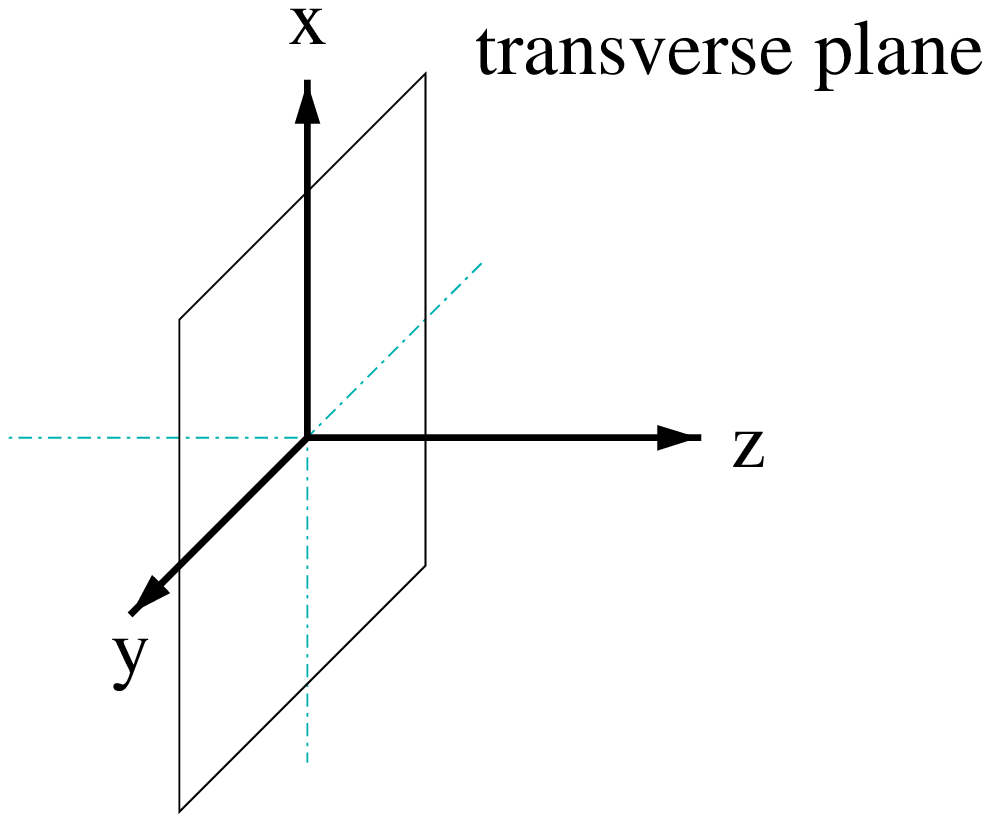, width=3cm}}}} &\nn \\
& \makebox[1cm][c]{$S_{LL}=$
    \raisebox{0.17cm}{\parbox[c]{5cm}{\epsfig{figure=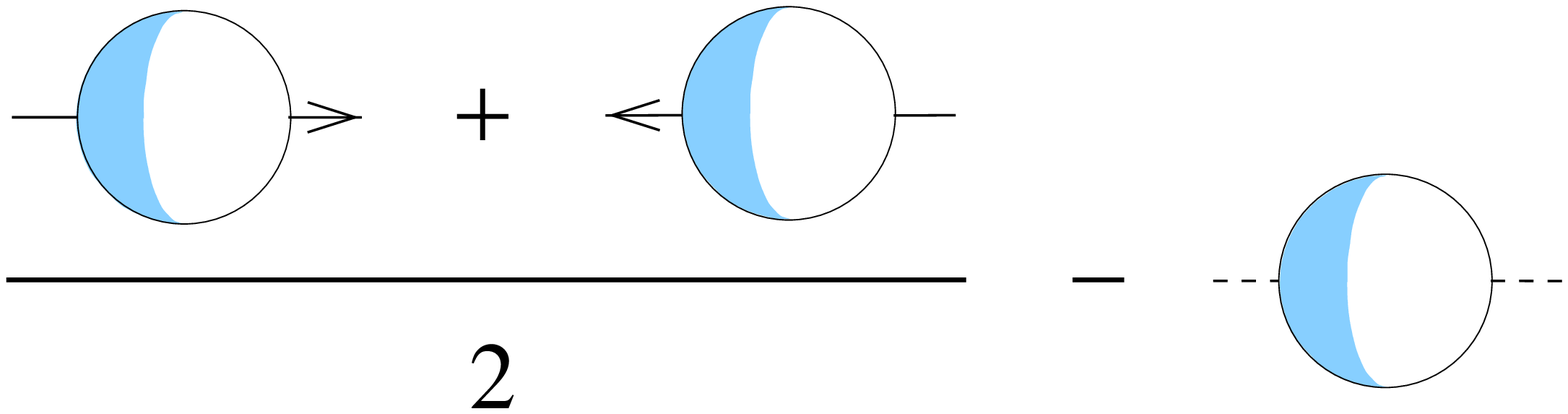, width=5cm}}}}& \\
S_{LT}^{x}=
   \makebox[4cm][c]{\parbox[c]{3.5cm}{\epsfig{figure=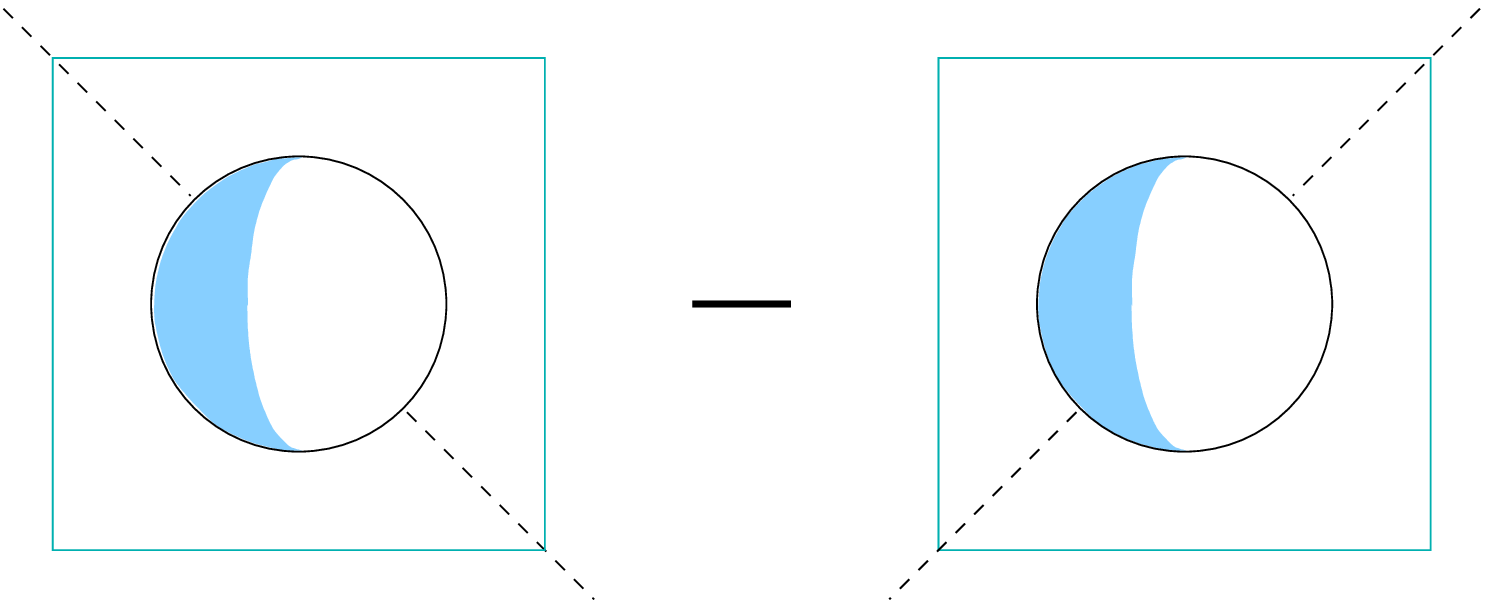, width=3.5cm}}} &&
S_{LT}^{y}=
	\parbox[c]{4cm}{\epsfig{figure=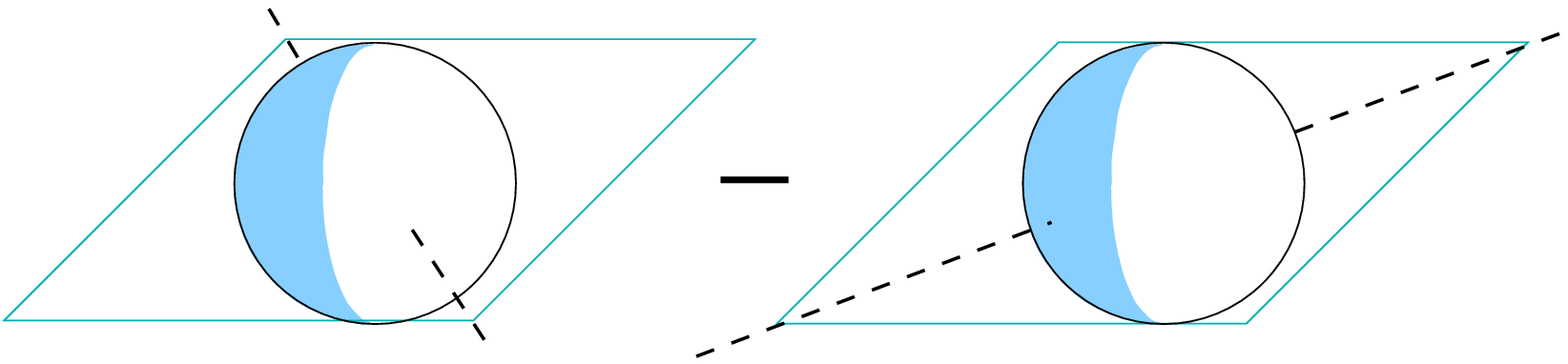, width=4cm}}	\\
S_{TT}^{xy}=
       \makebox[4cm][c]{\parbox[c]{3cm}{\epsfig{figure=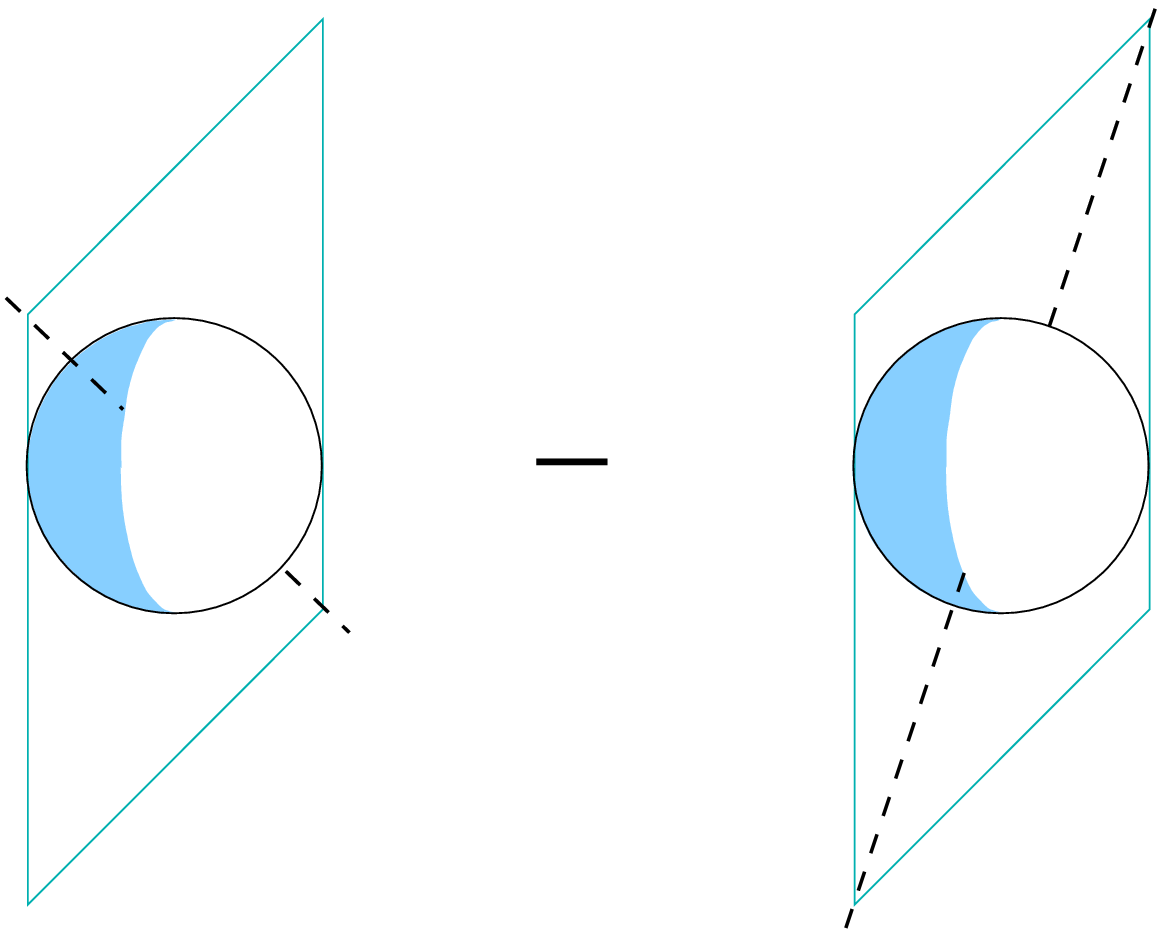, width=3cm}}}&& 
S_{TT}^{xx}=
	\makebox[4cm][c]{\parbox[c]{3cm}{\epsfig{figure=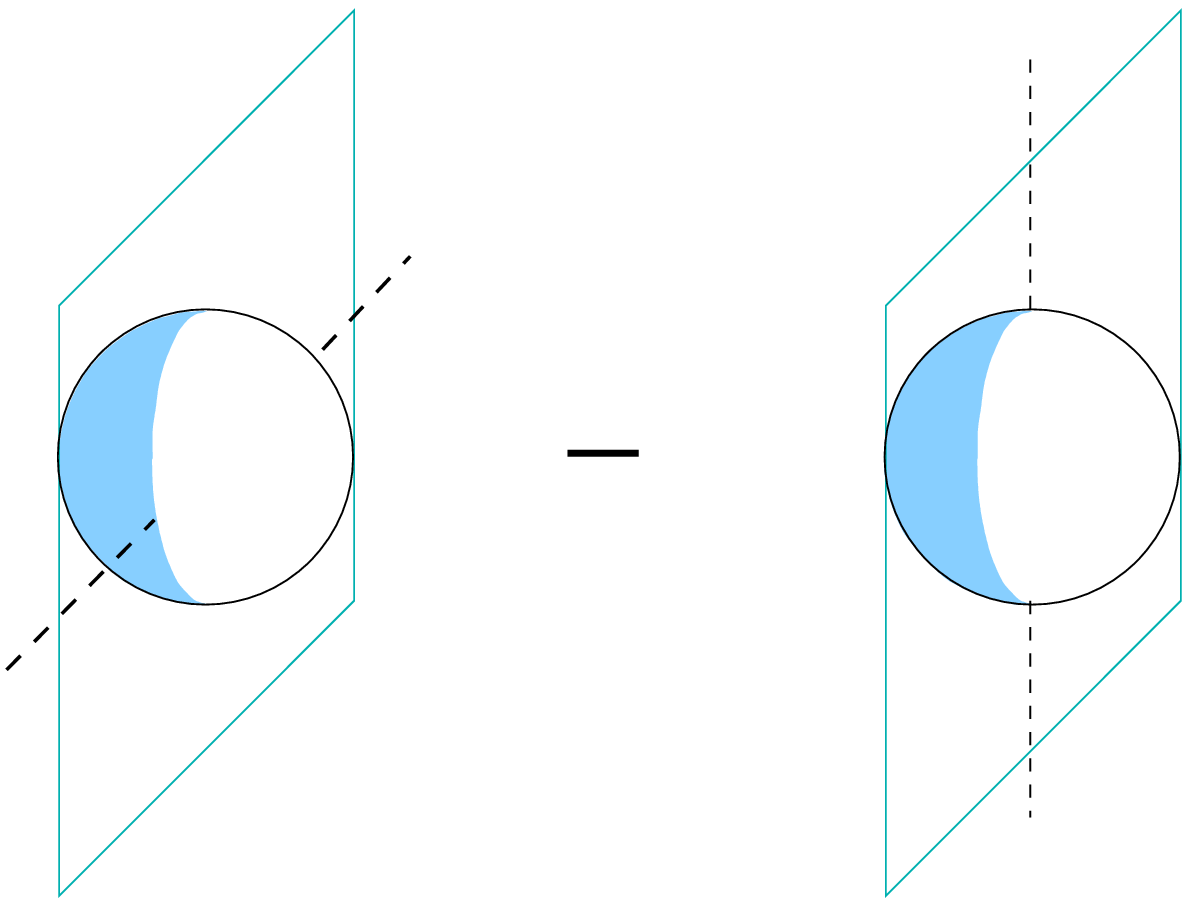, width=3cm}}}.
\end{eqnarray*} 

The probabilistic interpretations suggest straightforward  bounds 
on the values the spin tensor parameters can achieve, namely
\begin{equation}
\begin{array}{rcccl} 
-1 &\leq& S_{LL} &\leq& \frac{1}{2}, \\
-1 &\leq& S_{LT}^{i} &\leq& 1, \\
-1 &\leq& S_{TT}^{ij} &\leq& 1,
\end{array}
\end{equation}
where $i,j=x,y$.

Finally, it is possible to define a total degree of polarization
\begin{eqnarray} 
d &=& \left\{ \frac{3}{4}\, S^i S_i  
		+ \frac{3}{2}\, T^{ij} T_{ij} \right\}^{\frac{1}{2}} \nn \\
  &=& \left\{\frac{3}{4}\left[S_L^2 + (S_T^x)^2 + (S_T^y)^2 \right]\right.
\nn \\ && \qquad\left.\mbox{}
		+ \frac{3}{2}\left[\frac{2}{3}\,S_{LL}^2 
			+ \frac{1}{2}\left((S_{LT}^x)^2 +(S_{LT}^y)^2
			+(S_{TT}^{xx})^2+(S_{TT}^{xy})^2
			\right)\right]\right\}^{\frac{1}{2}},
\end{eqnarray}	
whose value ranges between 0 and 1.

\section{Measurement of spin tensor via decay analysis}
\label{a:decay}

In this appendix we show how it is possible to reconstruct the correspondence
between spin tensor and analyzing powers of
a $\rho$ meson by studying its decay into two pions.

In general the decay distribution of a spin-1 particle in two spin-0 particles
is given by
\begin{equation}
W(\theta,\varphi)= \tr \{ {\bm \rho}\,{\bm R}(\theta,\varphi) \},
\label{e:w}
\end{equation}
where $\theta$ and $\varphi$ are the polar and azimuthal angles of one of the
decay products in the parent particle's rest-frame.

The decay matrix $R$ is defined as
\begin{equation}
R_{mn}(\theta,\varphi)={\cal M}^{\dagger}_{m \,\rightarrow\, 0}(\theta,\varphi)
			  {\cal M}_{n \,\rightarrow\, 0}(\theta,\varphi).
\label{e:R1}
\end{equation}
The decay amplitudes can be written in terms of Wigner rotation
functions
\begin{equation}
\begin{array}{rcccl}
{\cal M}_{1 \,\rightarrow\, 0}(\theta,\varphi) &=&
	\sqrt{\frac{3}{4\pi}}\; D_{1 0}^{1\, *}(\varphi,\theta,-\varphi)
	&= &
		-\, \sqrt{\frac{3}{8\pi}}\, \sin{\theta}\; \e^{\ii \phi}, \\
{\cal M}_{0 \,\rightarrow\, 0}(\theta,\varphi) &=&
	\sqrt{\frac{3}{4\pi}}\; D_{0 0}^{1\, *}(\varphi,\theta,-\varphi) 
	&=&
		\sqrt{\frac{3}{4\pi}} \cos{\theta}, \\
{\cal M}_{-1 \,\rightarrow\, 0}(\theta,\varphi) &=&
	\sqrt{\frac{3}{4\pi}}\; D_{-1 0}^{1\, *}(\varphi,\theta,-\varphi) 
	&=&
		\sqrt{\frac{3}{8\pi}}\, \sin{\theta}\; \e^{- \ii \phi}. \\
\end{array}
\end{equation}
As it can be checked by explicit comparison, Eq.~(\ref{e:R1}) can be rewritten
as
\begin{equation}
{\bm R}(\theta,\varphi) = 
	\frac{1}{4 \pi} \left[\bm 1 + 3\, {\bm\Sigma}_{ij}
	\left(\frac{1}{3} \delta^{ij}
		-\^p_{\rm cm}^i(\theta,\varphi) \^p_{\rm cm}^j(\theta,\varphi)
	\right)\right],
\label{e:R2}
\end{equation}
where $\^p_{\rm cm}^i$ is the flight direction of one of the produced pions.

In general, the decay matrix can be expressed in terms of analyzing powers:
\begin{equation}
{\bm R}(\theta,\varphi) = \frac{1}{4 \pi}
	\left({\bm 1} + \frac{3}{2}\,{\bm \Sigma}_i\, A^i(\theta,\varphi)+
		 3\, {\bm\Sigma}_{ij}\, A^{ij}(\theta,\varphi) \right),
\label{e:R3}
\end{equation}
and the decay distribution can be obtained  accordingly as
\begin{equation}
W(\theta,\varphi)= \frac{1}{4 \pi}\left( {\bm 1} + \frac{3}{2}\,S_i\, A^i 
			+ 3\, T_{ij}\, A^{ij} \right)
\label{e:w2}
\end{equation}

By comparing Eq.~(\ref{e:R2}) with Eq.~(\ref{e:R3}) we can identify
\begin{eqnarray}
A^i&=&0 \nn \\
A^{ij} &=& \frac{1}{3} \delta^{ij}-\^p_{\rm cm}^i \^p_{\rm cm}^j. \nn
\end{eqnarray}

The tensor analyzing power can be written in a covariant form.  
By introducing the four-momenta of the two outgoing pions,
$P_{1}^{\mu}$ and $P_{2}^{\mu}$, since the two particles are identical we can 
make the replacement 
\begin{equation}
\^p^{\mu}_{\rm cm}=
	\frac{\left(P^{\mu}_{1}- P^{\mu}_{2}\right)}{|{\bm P}_1 - {\bm P}_2|}
	= \frac{\left(P^{\mu}_{1}- P^{\mu}_{2}\right)}
		{\sqrt{M_{\rho}^2 - 4 M_{\pi}^2}} 
\end{equation}
and we obtain the covariant expression of the tensor analyzing power
\begin{equation}
A^{\mu \nu} = \frac{1}{4 M_{\pi}^2 - M_{\rho}^2}
   \left(P^{\mu}_{1}- P^{\mu}_{2}\right)\left(P^{\nu}_{1}- P^{\nu}_{2}\right)
   -\frac{1}{3}\left(g^{\mu \nu} -\frac{P_h^{\mu}P_h^{\nu}}{M_{\rho}^2}\right).
\end{equation}

If the polar axis in the decay analysis is chosen along the $\rho$ direction of
motion, as it has been done in \cite{zeu,h1,her}, then we can use 
a parameterization for $A^{ij}$ analogous to that of the spin tensor,
Eq.~(\ref{e:tensor}), to obtain
\begin{equation}
\begin{array}{rclcrcl}
A_{LL}&=&-\frac{1}{2} \left(\cos^2{\theta}+\cos{2\theta}\right),\\
A_{LT}^x&=&-\sin{2\theta} \;\cos{\varphi}, &\qquad&
			 A_{LT}^y&=&-\sin{2\theta} \;\sin{\varphi},\\
A_{TT}^{xx}&=&-\sin^2{\theta}\; \cos{2\varphi}, &\qquad& 
			A_{TT}^{xy}&=&-\sin^2{\theta}\; \sin{2\varphi}.\\
\end{array}
\end{equation}

Substituting the explicit form of the decay matrix in Eq.~(\ref{e:w}), or
equivalently the explicit form of the tensor analyzing power 
in Eq.~(\ref{e:w2}), 
we obtain the decay distribution (cfr. \cite{sch})
\begin{eqnarray}
W( \theta, \varphi)
	&=& \frac{3}{8\pi}\left(\frac{2}{3}
	-\frac{2}{3}\,S_{LL}(\cos^2{\theta}+\cos{2\theta})
	-S_{LT}^{x} \sin{2\theta} \;\cos{\varphi} 
		-S_{LT}^{y} \sin{2\theta}\; \sin{\varphi} 
						\right. \nn \\ 
&&\left.\phantom{\frac{1}{1}}\mbox{}-S_{TT}^{xx}\sin^2{\theta}\; \cos{2\varphi}
	-S_{TT}^{xy}\sin^2{\theta}\; \sin{2\varphi}\right).
\end{eqnarray}

In case the polar axis is chosen in the direction of the virtual photon, 
in order to determine the relevant invariant quantity for $S_L$,
$S_{LL}$, $S_{LT}^\mu$ and $S_{TT}^{\mu\nu}$, we construct the
covariant comparison as in Eq.~(\ref{covtransf}), using the relation
between $g_\perp^{\mu\nu}$ and $g_T^{\mu\nu}$. It is then easy to
find for any hadron (neglecting order  $1/Q^2$ corrections),
\begin{eqnarray}
S_{L} &=& \frac{M\,(S\cdot q)}{P\cdot q},\\
S_{T}^\mu &=& S_{\perp}^\mu - S_{L}\,\frac{P_{\perp}^\mu}{M},\\
 \nonumber \\
\frac{2}{3}\,S_{LL} 
&=& \frac{M^2\,(q^\rho T_{\rho\sigma}q^\sigma)}{(P\cdot q)^2} ,\\ 
\frac{1}{2}\,S_{LT}^\mu 
&=& \frac{M\,(g_\perp^{\mu\rho}T_{\rho\sigma}q^\sigma)}{P\cdot q} -
\frac{2}{3}\,S_{LL}\,\frac{P_{\perp}^\mu}{M} ,\\ 
\frac{1}{2}\,S_{TT}^{\mu\nu} 
&=& g_\perp^{\mu\rho}T_{\rho\sigma}g_\perp^{\sigma\nu}
- \frac{1}{2}\,\frac{P_{\perp}^{\{\mu}S_{LT}^{\nu\}}}{M}
-\frac{2}{3}\,S_{LL}\,\frac{P_{\perp}^\mu P_{\perp}^\nu}{M^2} 
\nonumber \\
\mbox{}\hspace{1cm}
&=& g_\perp^{\mu\rho}T_{\rho\sigma}g_\perp^{\sigma\nu}
-\frac{P_{\perp}^{\{\mu}g_\perp^{\nu\}\rho}
T_{\rho\sigma}q^\sigma}{P\cdot q}
+ \frac{2}{3}\,S_{LL}\,\frac{P_{\perp}^\mu P_{\perp}^\nu}{M^2} .
\end{eqnarray}

\section{Distribution functions}
\label{a:distr}

Distribution functions can be defined in terms of 
projections of the correlation function on specific Dirac structures. Using 
 the notation
\begin{eqnarray} 
\Phi^{[\Gamma]}(x ,{\bm p}_T) &=& 
		\tr [\Phi(x ,{\bm p}_T)\, \Gamma], \\
\Phi^{[\Gamma]}(x) &=&
		\tr [\Phi(x)\, \Gamma], 
\end{eqnarray} 
we can list all the possible twist-2 projections and consequently define all
the possible twist-2 distribution function. In the following formulae 
distribution functions on 
the right side are understood to be functions of $x$ and $p_T^2$. Latin
indices, $i$, $j$ and $l$, indicate only the two transverse components. Before
integration upon ${\bm p}_T$ we obtain:
\begin{eqnarray}	
\Phi_U^{[\g^+]}(x, {\bm p}_T)& =& f_1, \nn \\
\Phi_L^{[\g^+]}(x, {\bm p}_T)& =& 0 , \nn \\
\Phi_T^{[\g^+]}(x, {\bm p}_T)& =&\left(\eps_T^{\mu \nu} S_{T\, \nu} 
		\frac{p_{T\,\mu}}{M}\, f_{1T}^{\perp}\right), \nn \\
\Phi_{LL}^{[\g^+]}(x, {\bm p}_T)& =&S_{LL}\, f_{1LL},  \\
\Phi_{LT}^{[\g^+]}(x, {\bm p}_T)& =&
		\frac{{\bm S}_{LT} \cdot {\bm p}_T}{M}\, f_{1LT}, \nn \\
\Phi_{TT}^{[\g^+]}(x, {\bm p}_T)& =&
		\frac{{\bm p}_T \cdot {\bm S}_{TT} \cdot {\bm p}_T}{M^2}
			\, f_{1TT}, \nn 
\end{eqnarray}
\begin{eqnarray}	
\Phi_U^{[\g^+ \g_5]}(x, {\bm p}_T)& =& 0, \nn \\
\Phi_L^{[\g^+ \g_5]}(x, {\bm p}_T)& =& S_L \, g_{1L}, \nn \\
\Phi_T^{[\g^+ \g_5]}(x, {\bm p}_T)& =&
		\frac{{\bm S}_T \cdot {\bm p}_T}{M}\, g_{1T}, \nn \\
\Phi_{LL}^{[\g^+ \g_5]}(x, {\bm p}_T)& =& 0,  \\
\Phi_{LT}^{[\g^+ \g_5]}(x, {\bm p}_T)& =&\left(\eps_T^{\mu \nu} S_{LT\,\nu}
		\frac{p_{T\,\mu}}{M}\, g_{1LT}\right), \nn \\
\Phi_{TT}^{[\g^+ \g_5]}(x, {\bm p}_T)& =&
	\left(-\eps_T^{\mu \nu}S_{TT\,\nu \rho}
			\frac{p_T^{\rho} p_{T\,\mu}}{M^2}\,g_{1TT}\right), \nn 
\end{eqnarray}
\begin{eqnarray}
\Phi_U^{[\ii \sig^{i+} \g_5]}(x, {\bm p}_T)& =& \left(
			\frac{\eps_T^{ij} p_{T\,j}}{M}
				\, h_1^{\perp}\right), \nn \\
\Phi_L^{[\ii \sig^{i+} \g_5]}(x, {\bm p}_T)& =& S_L \,
			\frac{p_T^i}{M}\,h_{1L}^{\perp} , \nn \\
\Phi_T^{[\ii \sig^{i+} \g_5]}(x, {\bm p}_T)& =&S_T^i\, h_{1T} +
			\frac{{\bm S}_T \cdot {\bm p}_T}{M} \frac{p_T^i}{M}\, 
			h_{1T}^{\perp}, \nn \\
\Phi_{LL}^{[\ii \sig^{i+} \g_5]}(x, {\bm p}_T)& =&\left( S_{LL} 
		   \frac{\eps_T^{ij}p_{T\,j}}{M}\, h_{1LL}^{\perp}\right),  \\
\Phi_{LT}^{[\ii \sig^{i+} \g_5]}(x, {\bm p}_T)& =& \left(
			\eps_T^{ij} S_{LT\,j}\, h'_{1LT}\right) +\left(
			\frac{{\bm S}_{LT} \cdot {\bm p}_T}{M} 
				\frac{ \eps_T^{ij}p_{T\,j}}{M}
				\, h_{1LT}^{\perp}\right), \nn \\
\Phi_{TT}^{[\ii \sig^{i+} \g_5]}(x, {\bm p}_T)& =& \left(
		        \eps_T^{ij} S_{TT\,jl}
			\frac{p_T^l}{M}\, h'_{1TT}\right) + \left(
		\frac{{\bm p}_T \cdot {\bm S}_{TT} \cdot {\bm p}_T}{M^2}
			\frac{\eps_T^{ij} p_{T\,j}}{M} 
			\,h_{1TT}^{\perp}\right). \nn 
\end{eqnarray}

After integrating over ${\bm p}_T$ the following distribution functions remain:
\begin{eqnarray}	
\Phi_U^{[\g^+]}(x)&=& f_1(x), \nn \\
\Phi_{LL}^{[\g^+]}(x)&=&S_{LL}\, f_{1LL}(x),   \nn \\
\Phi_L^{[\g^+ \g_5]}(x)&=& S_L \, g_{1}(x),  \\
\Phi_T^{[\ii \sig^{i+} \g_5]}(x)&=&S_T^i\, h_{1}(x) , \nn \\
\Phi_{LT}^{[\ii \sig^{i+} \g_5]}(x)&=& \left(
		\eps_T^{ij} S_{LT\,j}\, h_{1LT}(x)\right).  \nn
\end{eqnarray} 

The list of $p_T^\alpha$-weighted functions is
\begin{eqnarray}
\frac{1}{M}\left(\Phi_{\partial}^{\alpha}\right)_T^{[\g^+]} (x)& =&  
- \left(\eps_T^{\alpha \nu}S_{T\,\nu}\,f^{\perp (1)}_{1T}(x)\right), \\
\frac{1}{M}\left(\Phi_{\partial}^{\alpha}\right)_{LT}^{[\g^+]} (x)& =&  
-S_{LT}^{\alpha}\,f_{1LT}^{(1)}(x), \\
\frac{1}{M}\left(\Phi_{\partial}^{\alpha}\right)_T^{[\g^+ \g_5]} (x)& =&  
-S_T^{\alpha}\,g_{1T}^{(1)}(x), \\
\frac{1}{M}\left(\Phi_{\partial}^{\alpha}\right)_{LT}^{[\g^+ \g_5]} (x)& =&  
-\left(\eps_T^{\mu \alpha}S_{LT\,\mu}\,g_{1LT}^{(1)}(x) \right), \\
\frac{1}{M}\left(\Phi_{\partial}^{\alpha}\right)_U^{[\ii \sig^{i+} \g_5]} (x)
& =& 
-\left(\eps_T^{i \alpha}\,h_1^{\perp (1)}(x) \right),\\
\frac{1}{M}\left(\Phi_{\partial}^{\alpha}\right)_L^{[\ii \sig^{i+} \g_5]} (x)
& =&  
-S_L\,g_T^{i \alpha}\, h_{1L}^{\perp (1)}(x),\\
\frac{1}{M}\left(\Phi_{\partial}^{\alpha}\right)_{LL}^{[\ii \sig^{i+}\g_5]}(x)
& =&  
-\left(S_{LL} \,\eps_T^{i \alpha}\,h_{1LL}^{\perp (1)}(x)\right), \\
\frac{1}{M}\left(\Phi_{\partial}^{\alpha}\right)_{TT}^{[\ii \sig^{i+}\g_5]}(x)
& =& 
-\left(\eps_T^{i \mu}\,{S_{TT}}_{\mu}^{\alpha}\,h_{1TT}^{(1)}(x)\right),
\end{eqnarray}

The list of fragmentation functions can be obtained by applying 
 the notation replacements 
$f \rightarrow D$, $g \rightarrow G$,  $h \rightarrow H$ and the replacements 
$\{x,p_T,S,M,\g^+,\sig^{i+}\} \rightarrow  \{z,k_T,S_h,M_h,\g^-,\sig^{i-}\}$.

\section{Hadronic tensor with a tensor polarized outgoing fragment}
\label{a:hadronic}

We give the formulae for the complete hadronic tensor up to leading order in
1/Q and for different polarization conditions, starting from the expression
\begin{equation}
2 M W^{\mu \nu}=2z_h \int \de^2 {\bm p}_T \de^2 {\bm k}_T \,
	 \delta^2 ({\bm p}_T + {\bm q}_T - {\bm k}_T) 
	\;\tr \left[2 \Phi(\xbj,{\bm p}_T)\, \g^{\mu} 
		\; 2 \Delta(z_h,{\bm k}_T)\, \g^{\nu} \right].
\end{equation}

We limit ourselves to the case
where the target is a spin-$\frac{1}{2}$ hadron and the fragment is a spin-1
hadron (e.g. a $\rho$ meson whose polarization is
measured through its decay) with tensor polarization only. Therefore, spin
vector components refer to the target, while spin tensor 
components refer to the outgoing hadron (we label them with an index $h$).
When we use the expressions $S_{h\,LT}^{\mu}$ and $S_{h\,TT}^{\mu \nu}$ 
we mean the extensions to four dimension of the purely transverse vector 
${\bm S_{h\,LT}}$ and tensor ${\bm S_{h\,TT}}$. 
These extensions have therefore only transverse 
components.

\subsection{Unpolarized target -- tensor polarized fragment}

\begin{eqnarray} 
2M W_S^{\mu \nu}&=& 2 z \int \de^2 {\bm k}_T \de^2 {\bm p}_T \, 
			\delta^2\,({\bm p}_T +{\bm q}_T - {\bm k}_T)	\nn \\
	&& \mbox{}\times \left\{ 
	-g_{\perp}^{\mu \nu} \left[
		S_{h\,LL} \; f_1 \; D_{1LL} +
		\frac{{\bm S}_{h\,LT}\cdot {\bm k}_T}{M_h}
				\, f_1 \; D_{1LT} +
		\frac{{\bm k}_T \cdot {\bm S}_{h\,TT} \cdot {\bm k}_T}{M_h^2}
				\, f_1 \; D_{1TT} 
					\right] \right\}  \\
\nn \\ \nn \\
2M W_A^{\mu \nu}&=& 2 z \int \de^2 {\bm k}_T \de^2 {\bm p}_T \, 
			\delta^2\,({\bm p}_T +{\bm q}_T - {\bm k}_T)	\nn \\
	&& \mbox{}\times \left\{ 
	\ii \eps_{\perp}^{\mu \nu} \left[
	  \frac{{\bm k}_T \cdot \bold{\eps}_T \cdot {\bm S}_{h\,LT}}{M_h}
				\, f_1 \; G_{1LT} +
	\frac{({\bm k}_T \cdot \bold{\eps}_T)
			\cdot({\bm S}_{h\,TT} \cdot {\bm k}_T)}{M_h^2}
				\, f_1 \; G_{1TT} 
					\right]\right\}
\end{eqnarray}

\subsection{Longitudinally polarized target -- tensor polarized fragment}

\begin{eqnarray} 
2M W_S^{\mu \nu}&=& 2 z \int \de^2 {\bm k}_T \de^2 {\bm p}_T \, 
			\delta^2\,({\bm p}_T +{\bm q}_T - {\bm k}_T)	\nn \\
	&& \mbox{}\times \left\{ 
	-g_{\perp}^{\mu \nu} \left[
	S_L\;\frac{{\bm k}_T \cdot \bold{\eps}_T \cdot {\bm S}_{h\,LT}}{M_h}
				\, g_{1L} \; G_{1LT} 
	+S_L\;\frac{({\bm k}_T \cdot \bold{\eps}_T)\cdot({\bm S}_{h\,TT} 
			\cdot {\bm k}_T)}{M_h^2}
				\, g_{1L} \; G_{1TT}
					\right] \right.  \nn \\
	&& \mbox{}-
	\frac{p_{T}^{\{\mu}\eps_{\perp}^{\nu\} \tau}k_{T \tau}+
	k_{T}^{\{\mu}\eps_{\perp}^{\nu\} \tau}p_{T \tau}}{2MM_h} 
	\left[S_L\; S_{h\,LL} \; h_{1L}^{\perp} \; H_{1LL}^{\perp} +
		S_L\;\frac{{\bm S}_{h\,LT}\cdot {\bm k}}{M_h}
				\, h_{1L}^{\perp} \; H_{1LT}^{\perp}\right. 
	\nn \\ && \left.\makebox[6cm]{} +
	S_L\;\frac{{\bm k}_T \cdot{\bm S}_{h\,TT}\cdot {\bm k}_T}{M_h^2}
			\, h_{1L}^{\perp} \; H_{1TT}^{\perp}\right] \nn \\
	&& \mbox{}- 
	\frac{p_{T}^{\{\mu}\eps_{\perp}^{\nu\} \tau}S_{LT\,\tau}+
		S_{h\,LT}^{\{\mu}\eps_{\perp}^{\nu\} \tau}p_{T\,\tau}}{2M}
	\left[S_L\; h_{1L}^{\perp} H'_{1LT}\right] \nn \\
	&& \left.\mbox{}+ 
	\frac{p_{T}^{\{\mu}\eps_{\perp}^{\nu\} \tau}
				S_{TT\,\tau\sig}\,k_{T}^{\sig}+
	k_{T}^{\sig}\,S_{TT\,\sig}^{ \{\mu}\,\eps_{\perp}^{\nu\} \tau}
				\,p_{T}^{\tau}}{2MM_h}
	\left[S_L\; h_{1L}^{\perp} H_{1TT}\right] \right\}
 \\ \nn \\
2M W_A^{\mu \nu}&=& 2 z \int \de^2 {\bm k}_T \de^2 {\bm p}_T \, 
			\delta^2\,({\bm p}_T +{\bm q}_T - {\bm k}_T)	\nn \\
	&& \mbox{}\times \left\{ 
	\ii \eps_{\perp}^{\mu \nu} \left[
		S_L\; S_{h\,LL} \; g_{1L} \; D_{1LL} +
		S_L\;\frac{{\bm S}_{h\,LT}\cdot {\bm k}_T}{M_h}
				\, g_{1L} \; D_{1LT} \right.\right.
\nn \\&&\left.\makebox[6cm]{}
	\left.\mbox{}+S_L\;\frac{{\bm k}_T \cdot {\bm S}_{h\,TT} 
				\cdot {\bm k}_T}{M_h^2}
				\, g_{1L} \; D_{1TT}
					\right]\right\} 
\end{eqnarray}

\subsection{Transversely polarized target -- tensor polarized fragment}

\begin{eqnarray} 
2M W_S^{\mu \nu}&=& 2 z \int \de^2 {\bm k}_T \de^2 {\bm p}_T \, 
			\delta^2\,({\bm p}_T +{\bm q}_T - {\bm k}_T)	\nn \\
	&& \mbox{}\times \left\{ 
	-g_{\perp}^{\mu \nu} \left[
	\frac{{\bm S}_T\cdot {\bm p}_T}{M}\;
	\frac{{\bm k}_T \cdot \bold{\eps}_T \cdot {\bm S}_{h\,LT}}{M_h}
				\, g_{1T} \; G_{1LT} 
	+\frac{{\bm S}_T\cdot {\bm p}_T}{M}\;
		\frac{({\bm k}_T \cdot \bold{\eps}_T)
			\cdot({\bm S}_{h\,TT} \cdot {\bm k}_T)}{M_h^2}
				\, g_{1T} \; G_{1TT}
					\right]\right.  \nn \\
&& \mbox{}- 
	\frac{p_{T}^{\{\mu}\eps_{\perp}^{\nu\} \tau}k_{T \tau}+
	k_{T}^{\{\mu}\eps_{\perp}^{\nu\} \tau}p_{T \tau}}{2MM_h} 
	\left[\frac{{\bm S}_T\cdot {\bm p}_T}{M}\; S_{h\,LL} \; h_{1T}^{\perp} 
				\; H_{1LL}^{\perp} +
   \frac{{\bm S}_T\cdot {\bm p}_T}{M}\;\frac{{\bm S}_{h\,LT}\cdot {\bm k}}{M_h}
				\, h_{1T}^{\perp} \; H_{1LT}^{\perp}\right. 
	\nn \\ 
       && \left.\makebox[6cm]{} +
	\frac{{\bm S}_T\cdot {\bm p}_T}{M}\;
		\frac{{\bm k}_T \cdot{\bm S}_{h\,TT}\cdot {\bm k}_T}{M_h^2}
			\, h_{1T}^{\perp} \; H_{1TT}^{\perp}\right]  \nn \\
&& \mbox{}- 
	\frac{S_{T}^{\{\mu}\eps_{\perp}^{\nu\} \tau}k_{T \tau}+
	k_{T}^{\{\mu}\eps_{\perp}^{\nu\} \tau}S_{\perp \tau}}{2M_h}
	\left[ S_{h\,LL} \; h_{1T} \; H_{1LL}^{\perp} +
       \frac{{\bm S}_{h\,LT}\cdot {\bm k}}{M_h}
				\, h_{1T} \; H_{1LT}^{\perp}\right. 
	\nn \\ && \left.\makebox[6cm]{} +
	\frac{{\bm k}_T \cdot{\bm S}_{h\,TT}\cdot {\bm k}_T}{M_h^2}
			\, h_{1T} \; H_{1TT}^{\perp}\right]  \nn \\ 
&& \mbox{}- 
	\frac{p_{T}^{\{\mu}\eps_{\perp}^{\nu\} \tau}S_{LT\,\tau}+
		S_{h\,LT}^{\{\mu}\eps_{\perp}^{\nu\} \tau}p_{T\,\tau}}{2M}
	\left[\frac{{\bm S}_T\cdot {\bm p}_T}{M}\; 
				h_{1T}^{\perp} H'_{1LT}\right] \nn \\
&& \mbox{}+
	\frac{p_{T}^{\{\mu}\eps_{\perp}^{\nu\} \tau}
				S_{TT\,\tau\sig}\,k_{T}^{\sig}+
	k_{T}^{\sig}\,S_{TT\,\sig}^{ \{\mu}\,\eps_{\perp}^{\nu\} \tau}
				\,p_{T}^{\tau}}{2MM_h}
	\left[\frac{{\bm S}_T\cdot {\bm p}_T}{M}\;
				h_{1T}^{\perp} H_{1TT}\right] \nn \\ 
&& \mbox{}- \frac{S_{T}^{\{\mu}\eps_{\perp}^{\nu\} \tau}S_{LT\,\tau}+
       S_{h\,LT}^{\{\mu}\eps_{\perp}^{\nu\} \tau}S_{\perp\,\tau}}{2}
	\left[h_{1T} H'_{1LT}\right] \nn \\
&& \left.\mbox{}+ 
	\frac{S_{T}^{\{\mu}\eps_{\perp}^{\nu\} \tau}
				S_{TT\,\tau\sig}\,k_{T}^{\sig}+
	k_{T}^{\sig}\,S_{TT\,\sig}^{ \{\mu}\,\eps_{\perp}^{\nu\} \tau}
				\,S_{T}^{\tau}}{2M_h}
	\left[h_{1T} H_{1TT}\right] \right\}
 \\ \nn \\
2M W_A^{\mu \nu}&=& 2 z \int \de^2 {\bm k}_T \de^2 {\bm p}_T \, 
			\delta^2\,({\bm p}_T +{\bm q}_T - {\bm k}_T)	\nn \\
	&& \mbox{}\times \left\{ 
	\ii \eps_{\perp}^{\mu \nu} \left[
	    \frac{{\bm S}_T\cdot {\bm p}_T}{M}\; S_{h\,LL}\;g_{1T} \; D_{1LL} +
		\frac{{\bm S}_T\cdot {\bm p}_T}{M}\;
			\frac{{\bm S}_{h\,LT}\cdot {\bm k}_T}{M_h}
				\, g_{1T} \; D_{1LT}\right. \right.
	\nn \\ && \left. \left.\makebox[6cm]{} +
   \frac{{\bm S}_T\cdot {\bm p}_T}{M}\;\frac{{\bm k}_T 
			\cdot {\bm S}_{h\,TT} \cdot {\bm k}_T}{M_h^2}
				\, g_{1T} \; D_{1TT}
					\right]\right\}. 
\end{eqnarray}


\end{document}